\documentclass[a4paper,11pt]{article}

\usepackage{subcaption}
\usepackage{standalone}
\usepackage{pgfplots}
\pgfplotsset{compat=1.16}
\usepackage{jheppub} %
\usepackage{physics}
\usepackage{derivative}

\usepackage{amssymb}
\usepackage{amsmath}
\usepackage{iftex}
\ifLuaTeX
\usepackage{fontspec} %
\else
\usepackage[utf8]{inputenc} %
\usepackage[T1]{fontenc} %
\fi

\usepackage{float}
\usepackage{xcolor}
\usepackage{cancel}
\usepackage{bm}
\usepackage{enumitem} %
\usepackage{mathtools}
\usepackage{braket}
\usepackage{mdframed}
\usepackage{cleveref}
\usepackage{ytableau}
\usepackage{comment}
\usepackage{amstext} %
\usepackage{lineno}
\usepackage{nicematrix}
\usepackage{xspace} %

\usepackage{slashed}

\def\det{\mathop{\rm det}\nolimits}

\def\Tr{\mathop{\rm Tr}\nolimits}
\def\bbra{{\langle\kern-2.5pt\langle}}
\def\kket{{\rangle\kern-2.5pt\rangle}}
\def\Bbra{{\Big\langle\kern-3.5pt\Big\langle}}
\def\Kket{{\Big\rangle\kern-3.5pt\Big\rangle}}

\newcommand   \half{\frac 1 2}
\newcommand   \lptl{\raise .8ex\hbox{$^\leftarrow$} \hspace{-9pt} \partial}
\newcommand   \lrptl{\raise .8ex\hbox{$^\leftrightarrow$} \hspace{-9pt} \partial}

\renewcommand \L  {\Lambda}

\newcommand   \SO    {\mathrm{SO}}
\newcommand   \gO    {\mathrm{O}}
\newcommand   \gU    {\mathrm{U}}

\newcommand   \Sp    {\mathrm{Sp}}

\DeclareMathOperator{\vol}{vol}

\newcommand   \cD {\mathcal{D}}

\newcommand   \cF {\mathcal{F}}

\newcommand   \cL {\mathcal{L}}

\newcommand   \cN {\mathcal{N}}
\newcommand   \cO {\mathcal{O}}

\newcommand   \cZ {\mathcal{Z}}

\newcommand{\be}{
  \begin{equation}
  \begin{aligned}
}
\newcommand{\ee}{
  \end{aligned}
  \end{equation}
}

\makeatletter
\DeclareRobustCommand\bea{\@ifnextchar[{\@@bea}{\@bea}}
\def\@@bea[#1]#2\eea{\begin{subequations}\begin{align}#2\end{align}\label{#1}\end{subequations}}
\def\@bea#1\eea{\begin{subequations}\begin{align}#1\end{align}\end{subequations}}
\makeatother

\usepackage[compat=1.1.0]{tikz-feynman}
\tikzfeynmanset{warn luatex=false}

\usepackage{array}   %
\newcolumntype{L}{>{$}l<{$}} %
\makeatletter
\edef\savedcodes{\catcode`\noexpand\_=\the\catcode`\_}
\edef\@tempa{\csname opt@newtxmath.sty\endcsname}
\def\@tempb{{subscriptcorrection}}
\expandafter\expandafter\expandafter
  \in@\expandafter\@tempb\expandafter{\@tempa}
\ifin@
  \catcode`\_=12 %
\fi
\makeatother
\usepackage{tensor}
\savedcodes

\newcommand{\verticalcenter}[1]{\vcenter{\hbox{#1}}}

\title{$F$-extremization determines certain large-$N$ CFTs}

\author[a,1]{Ludo Fraser-Taliente,\note{Corresponding author.}}

\author[a]{and John Wheater}

\affiliation[a]{Rudolf Peierls Centre for Theoretical Physics, University of Oxford, Oxford OX1 3PU, UK}

\emailAdd{ludovic.fraser-taliente@physics.ox.ac.uk}

\abstract{
We show that the conformal data of a range of large-$N$ CFTs, the melonic CFTs, are specified by constrained extremization of the universal part of the sphere free energy $F=-\log Z_{S^d}$. 
This family includes the generalized SYK models, the vector models (O$(N)$, Gross-Neveu, etc.), and the tensor field theories. %
The known $F$ and $a$-maximization procedures in SCFTs are therefore extended to these non-supersymmetric CFTs in continuous $d$.
We establish our result using the two-particle irreducible (2PI) effective action, and, equivalently, by Feynman diagram resummation.
The universal part of $F$ interpolates in continuous dimension between the known $C$-functions, so we can interpret this result as an extremization of the number of IR degrees of freedom, in the spirit of the generalized $c,F,a$-theorems.
The outcome is a complete classification of the melonic CFTs: they are the conformal mean field theories which extremize the universal part of the sphere free energy, subject to an IR marginality condition on the interaction Lagrangian. 
}

\newcommand{\id}{\mathbb{I}}
\newcommand{\spinid}{\mathbb{I}_s}
\newcommand{\thalf}{\tfrac{1}{2}}
\DeclareMathOperator{\Str}{\mathrm{Str}}
{%
}%

\newcommand{\Gglobal}{\mathcal{G}}
\newcommand{\rhoext}{\rho^\prime}
\newcommand{\Ft}{\tilde{F}}
\newcommand{\FttextOrPDF}{\texorpdfstring{$\Ft$}{F~}\xspace}
\newcommand{\FtextOrPDF}{\texorpdfstring{$F$}{F}\xspace}

\newcommand{\twoPt}{D}

\usepackage{empheq} %
\newcommand*\widefbox[1]{\fbox{\hspace{2em}#1\hspace{2em}}}

\begin{document}
\maketitle
\flushbottom

\section{Introduction}

In this paper, we show that the conformal data of a particular family of large-$N$ CFTs, the melonic CFTs, are determined by constrained extremization of $\Ft$, the universal part of the sphere free energy of a collection of generalized free fields.
The constraints arise directly from the interaction terms, and are linear in the conformal scaling dimensions of the fields. Put another way, the melonic CFTs are precisely the conformal mean field theories with constrained extremal $\Ft$. Notably, this procedure turns out to be identical to the $F$ and $a$-maximization principles used to determine the $R$-charges and scaling dimensions of SCFTs with four supercharges \cite{Giombi:2014xxa, Pufu:2016zxm}.

The melonic CFTs arise as the conformal vacua of the melonic quantum field theories, which are a family of large-$N$ QFTs that have a resummable diagrammatic expansion. Recently discovered, they have a structure that is simpler than that of the matrix models, but nonetheless lead to exactly solvable large-$N$ CFTs \cite{Benedetti:2020seh}.
It is useful to distinguish three principal types of melonic QFTs, all of which occur in the strict large-$N$ limit:
\begin{itemize} 
    \item the Sachdev-Ye-Kitaev (SYK) model and its generalizations \cite{Gross:2016kjj};
    \item the tensor models $\phi_{a_1\cdots a_r}$ with $\gO(M)^r$ symmetry for rank $r\ge 3$  and $M^r=N$ \cite{Witten:2016iux,Giombi:2017dtl};
    \item the vector models, such as the $\gO(N)$ $\phi^4$ model \cite{zinn-justin_quantum_2002}. 
\end{itemize}
The solvability of each of these models arises from the exact resummation of their Feynman-diagrammatic expansions at leading order in $N$. Using this, all the known melonic CFTs can and have been solved individually, whether in the SYK-like \cite{Maldacena:2016hyu,Rosenhaus:2018dtp,Murugan:2017eto,Liu:2018jhs,Bulycheva:2017ilt,Yoon:2017nig,Bulycheva:2017uqj,Gu:2019jub,Turiaci:2017zwd, Marcus:2018tsr,Fu:2016vas,Berkooz:2017efq,Chang:2021fmd,Chang:2023gow,Biggs:2023mfn}, tensor \cite{Witten:2016iux, Choudhury:2017tax, Klebanov:2018fzb, Gurau:2019qag, Gubser:2018yec,Fraser-Taliente:2024rql, Prakash:2017hwq, Klebanov:2019jup,Benedetti:2019rja,Chang:2018sve}, vector \cite{zinn-justin_quantum_2002,Chang:2021wbx}, or other \cite{Benedetti:2020iku} cases.  The resummability occurs for a slightly different (albeit related) reason in each case: for SYK-like theories, a disorder average over the coupling; for the vector and tensor models, the tuned combinatorics. However, for our purposes, the particular mechanism used is irrelevant -- the solvable CFT found in the IR is not sensitive to those details\footnote{Of course, the $N$-subleading dynamics of these mechanisms will certainly differ \cite{Gurau:2019qag,Choudhury:2017tax}.}. 

In this paper, we show that for any melonic QFT$_d$, regardless of the individual complexities of the model in question, the IR CFT$_d$ is determined by a universal principle: extremization of $\Ft$, where $\Ft$ is defined for a mean field theory with the same field content as the QFT but arbitrary conformal scaling dimensions. This reflects our expectation that in the large-$N$ limit, factorization means that the leading order in $N$ is essentially Gaussian, i.e. mean field. Since $\Ft$ is thought to count the effective number of degrees of freedom of a CFT$_d$ \cite{Giombi:2014xxa,Fei:2015oha} (being a candidate weak $C$-function in the sense of Zamolodchikov \cite{Zamolodchikov:1986c,Cardy:1988cwa}), this has an appealing simplicity.

In outline, the $\Ft$-extremization procedure is as follows. In arbitrary dimension $d$, we define a UV theory of order $\sim N$ free fields $\{\phi\}$ in arbitrary Lorentz and global symmetry representations. 
We perturb by a particular relevant interaction and follow the renormalization group flow to the deep IR, where we reach the melonic CFT of interest. There, the conformal symmetry means that the fields have some conformal scaling dimensions $\Delta_\phi$.
Thanks to the simplifications of the melonic limit, we can compute the universal part of the sphere free energy as a function of the unknown $\Delta_\phi$s,
\begin{equation}\begin{aligned}
\Ft \equiv \sum_{\text{fields } \phi} \Ft_{\phi}(\Delta_{\phi}). \label{eq:melonicFtSum}
\end{aligned}\end{equation}
This is interpreted as $\Ft$ for a mean field theory that has the same field content as the original theory, but each field $\phi$ has an arbitrary \textit{trial} dimension $\Delta_\phi$. Then, the \textit{actual} IR scaling dimensions are precisely those that extremize this $\Ft(\{\Delta_\phi\})$, subject only to the constraint that the potential evaluated in the IR is marginal (i.e. of dimension $d$). 
Generically, this procedure leads to a finite number of vacua in the IR for each $d$. 
Explicitly, for a perturbing melonic interaction of schematic form\footnote{We have suppressed the details that ensure the melonic resummability.}
\begin{equation}\begin{aligned}
S_{\mathrm{int}} \supset \sum_m g_m\int \dd^d x \prod_{\text{fields }\phi} \phi^{q_{\phi}^m},
\end{aligned}\end{equation}
the constraints are IR marginality of the integrated operator that each $g_m$ multiplies:
\begin{equation}\begin{aligned}
[\dd^d x] +\sum_{\text{fields } \phi} q^m_{\phi} \Delta_{\phi} = 0, \quad [\dd^d x]=-d; \label{eq:melonicConstraints}
\end{aligned}\end{equation} 
we call these the \textit{melonic constraints}. However, we must allow for the possibility that any given $g_m$ runs to zero, in which case the associated constraint is not applied.
Remarkably, this formulation makes manifest that any apparent dependence on the details of the global symmetry representations, or any particular complicated form for the interaction (especially for non-scalar Lorentz representations) is washed out by the conformal melonic limit. The only pieces of data that matter are:
\begin{enumerate}
    \item the monomial form of the relevant melonic interactions (given by the integers $\{q^m_\phi\}$);
    \item and the dimensions of the global symmetry representations of the fields.
\end{enumerate}
In the conformal large-$N$ limit, any non-Gaussian behaviour in the correlation functions of the CFT is subleading in $N$.
Therefore, the melonic CFTs are the conformal mean field theories, with some given field content, that extremize $\Ft$, subject to the constraints \eqref{eq:melonicConstraints}.

Because the universal part of the sphere free energy ($\Ft$) interpolates between the Weyl anomaly in even $d$ and the free energy in odd $d$, this procedure is identical to the mechanisms of $a$- and $F$-extremization in superconformal field theories (SCFTs) with four supercharges, without the large-$N$ limit\footnote{If the SCFTs are unitary, it can be shown that the extrema are actually maxima.} \cite{Giombi:2014xxa}: %
we need only swap out the $\phi$s for chiral superfields $\Phi$, and space for superspace.
The constraints \eqref{eq:melonicConstraints} are then exactly the SUSY-preserving constraints in the IR. A melonic version of such an SCFT can be solved using either the supersymmetry or the melonicity; indeed, the two procedures are identical in the SUSY-preserving sector. However, the melonic procedure extends to SUSY-breaking vacua.

The $\Ft$-extremization procedure is proved using the two-particle-irreducible (2PI) effective action \cite{Benedetti:2018goh}, or, equivalently, using the Schwinger-Dyson equations. In this paper, we give both proofs. The effective action approach is more enlightening, and we focus on it: in particular, $\Ft(\{\Delta_\phi\})$ \textit{is} the 2PI effective action evaluated with the conformal ansatz on the sphere. At its extremum, this function coincides with the universal part of the CFT sphere free energy, which is the usual quantity $\Ft_\mathrm{CFT}=\sin(\pi d/2)\log Z_{S^d}$ \cite{Giombi:2014xxa,Fei:2015oha}. %

The proof reduces to two components. The first is straightforward: since the full quantum solution functionally extremizes the 2PI effective action, the non-perturbative conformal scaling dimensions must also extremize $\Ft$. The second component is then demonstrating that in the conformal limit the running coupling constants become Lagrange multipliers implementing the constraints \eqref{eq:melonicConstraints}.
There are no further contributions, so the effective action becomes simply the sum of the sphere free energies of generalized free fields, with a constraint on their scaling dimensions, which must vanish at the physical point.
This gives a concrete understanding of why the $\Ft$-extremization procedure works. 
Crucially, however, the proof that the interaction contributes only linear constraints on the IR scaling dimensions rests on the particular properties of the melonic limit; except for the supersymmetric melonic case mentioned above, this structure will not persist to subleading orders in $N$.

We begin in \cref{sec:MelonicFTs} with the definition of the melonic field theories.
In \cref{sec:FthmsFmax} we review the role of the universal part of the sphere free energy in QFT, generalized free fields, and the various maximization procedures appearing in supersymmetric quantum field theories.
We give a careful explanation of our claim of $\Ft$-extremization in \cref{sec:fundamentalClaim}, and illustrate it with a simple example.
Our claim is proved in \cref{sec:FtMaxFrom2PI}, using the 2PI effective action (\cref{app:diagrammaticProof} contains an alternative Feynman-diagrammatic proof using the Schwinger-Dyson equations).
In the remainder of the paper, we explore the IR structure of these melonic CFTs as a function of $d$.
Much of this structure can be understood by comparison with the more familiar critical vector models, which we examine from an $\Ft$-extremization point of view in \cref{sec:vectorModels}; in \cref{sec:melonicModels}, we consider as examples three melonic CFTs, one of which is an SCFT, and outline the similarities.
We conclude with an outlook in \cref{sec:Discussion}.

\newcommand{\fr}{\mathfrak{r}}
\newcommand{\nrows}[1]{r(#1)}

\section{Melonic field theories and their IR limits} \label{sec:MelonicFTs}

\begin{figure}
    \centering
    \includegraphics[width=0.8\textwidth]{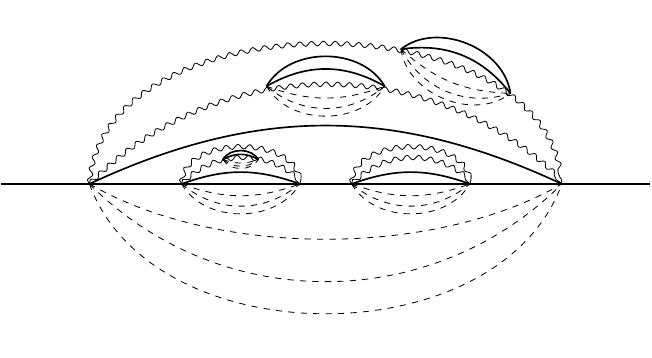}
    \caption{Iterated insertions on a melonic contribution to a two-point function of a theory containing three fields, with $q_i = \{3,2,2\}$.}
    \label{fig:recursiveMelons}
\end{figure}

The melonic QFTs are found in the large-$N$ limit of QFTs with interaction terms of schematic form
\begin{equation}\begin{aligned}
g_m \prod_{\text{fields }\phi} \phi^{q^m_{\phi}}, \quad q^m_{\phi} \in \mathbb{N}_0,
\end{aligned}\end{equation} 
where we have suppressed any index structure. These theories are then melonic if they possess a Feynman-diagrammatic expansion of their two-point functions $\expval{\Phi(x) \Phi(y)}$ that, at leading order in the large-$N$ limit, is dominated by diagrams of melonic form \cite{Benedetti:2023mli}
\begin{equation}\begin{aligned}
\Pi_{\phi} \supset  \Phi \vcenter{\hbox{\begin{tikzpicture}[anchor=base,baseline]
  \pgfmathsetmacro{\L}{4}
  \coordinate (L) at (-\L,0);
  \coordinate (R) at (\L,0);
  
  \draw[black, decorate, decoration={snake, amplitude=.4mm, segment length=2mm}]  (L) to[bend left=90,looseness=1.25] (R);
  \draw[black, decorate, decoration={snake, amplitude=.4mm, segment length=2mm}]  (L) to[bend left=75,looseness=1.20] (R);
  \draw[loosely dotted, thick] (0, {0.65*\L}) to (0, {0.55*\L});
  \node at (0.4, {(0.58*\L)}) {$q^m_{\phi_i}$};
  \draw[black, decorate, decoration={snake, amplitude=.4mm, segment length=2mm}]  (L) to[bend left=60] (R);
  
  \draw[black, thick] (L) to[bend left=50] (R);
  \draw[black, thick] (L) to[bend left=40, looseness=1] (R);
  \draw[loosely dotted, thick] (0, {0.32*\L}) to (0, {0.22*\L});
  \node at (0.8, {(0.25*\L)}) {$(q^m_{\Phi} -1)$};
  \draw[black, thick] (L) to[bend left=25, looseness=0.7] (R);

  \draw[black, thick] (-\L-1.5, 0) to (L);
  \draw[loosely dotted, thick] (0, {0.09*\L}) to (0, {-0.09*\L});
  \draw[black, thick] (R) to (\L+1.5,0);
  
  \draw[black, dashed] (L) to[bend right=40, looseness=1] (R);
  \draw[loosely dotted, thick] (0, {-0.23*\L}) to (0, {-0.33*\L});
  \node at (0.4, {-(0.29*\L)}) {$q^m_{\phi_k}$};
  \draw[black, dashed] (L) to[bend right=25, looseness=0.7] (R);
  \draw[black, dashed] (L) to[bend right=70] (R);
  \draw[black, dashed]  (L) to[bend right=90,looseness=1.25] (R);
\end{tikzpicture}}} \Phi. 
\end{aligned}\end{equation} 
We refer to this diagram as the melon diagram, for obvious reasons.
The full self-energy $\Pi_{\Phi}$ of the field $\Phi$, is then given by the melon diagram, and all possible nested re-insertions of the melon. That is, each $\phi_j$ propagator within the melon can have a recursive insertion of a melon with $q^m_{\phi_k}$ $\phi_k$ propagators for $k\neq j$, and $q^m_{\phi_j}-1$ $\phi_j$ propagators, as illustrated in \cref{fig:recursiveMelons}.
The sum of all of these \textit{melonic diagrams} gives a geometric series, which is resummable. Mathematically, in the case of the tensor models, these dominant diagrams are those of leading \textit{Gurau degree} \cite{Gurau:2019qag,Benedetti:2023mli}. %

The melonic diagrams are a summable subset of the planar diagrams, and therefore lie between the vector models and matrix models in terms of complexity: above the vector models, because they do not have the ultralocal dynamics of the vector models; but below the matrix models, since the melonic diagrams can in fact be resummed. Therefore, the data that defines a melonic field theory is simply the collection of integers $q^m_\phi$, which determine which melons contribute to the resummation of each $\phi$ propagator.
The result of this paper is that once a list of fields is given, the conformal IR solution is completely specified by this set of integers $q^m_{\phi}$, up to a discrete choice of vacua.

The canonical theories in the melonic class are those containing only a single field:
\begin{itemize}
    \item The $\psi^q$ SYK model is defined by $q_{\psi} = q$ \cite{Kitaev:2017awl,Maldacena:2016hyu}.
    \item The $\phi^q$ tensor model is defined by $q_{\phi}=q$ \cite{Witten:2016iux,Giombi:2017dtl,Carrozza:2018psc,Benedetti:2019rja,Harribey:2021xgh}.
\end{itemize}
The \textit{multi-field} melonic models, i.e. those with multiple distinct fundamental fields, display much richer IR structure, because the melonic constraints \eqref{eq:melonicConstraints} no longer completely solve the theory:
\begin{itemize}
    \item The generalized SYK model is defined \cite{Gross:2016kjj} by $q_{\psi_i} = q_i$ (hence our choice of notation).
    \item We can also consider multi-field tensor models, as in \cite{Giombi:2018qgp,Prakash:2022gvb,Fraser-Taliente:2024rql}, with $\gO(M)^r$ or indeed $\Sp(M)^r$ symmetry \cite{Gurau:2022dbx} ($M^r = N$).
    \item The supersymmetric SYKs or tensor models can be considered in two different ways:
    \begin{itemize}
        \item They can be formulated directly in superspace, using superfields $\Phi$. In that case, they correspond to single-field melonic models. Taking a superpotential of the form $W\sim \Phi^q$, we have simply $q_{\Phi}=q$ \cite{Chang:2018sve, Popov:2019nja, Lettera:2020uay}.
        \item They can also be formulated in components, $\Phi= \phi + \theta \psi - \theta^2 X$, in which case we obtain the simplest example of a multi-interaction melonic theory. A superpotential $W\sim \Phi^q$ (see (3.3) of \cite{Chang:2018sve}) decomposes into the two melonic interactions
    \begin{equation}\begin{aligned}
       g_1: \, \{q^1_\phi = q-2, \, q^1_\psi = 2, q^1_F =0\}, \quad g_2: \,
        \{q^2_\phi = q-1, \, q^2_\psi = 0, q^2_F =1\}.
    \end{aligned}\end{equation} 
    Component form permits the consideration of supersymmetry-breaking dynamics; we expand upon this in \cref{sec:SUSYmelons}.
    \end{itemize}
    \item The leading order of the large-$n$ $O(n)$ $\phi^4$ vector model, rewritten with an auxiliary field $\sigma=\phi_I\phi_I$, is defined by $q_{\sigma}=1$ and $q_{\phi} =2$. The same holds for the large-$n$ $U(n)$ Gross-Neveu model. We elaborate on this in \cref{sec:vectorModels}.
\end{itemize}
There are also further melonic mechanisms: one such is the Amit-Rogansky model. This is the $d$-dimensional theory of $N$ scalar fields in an irrep of $\SO(3)$, with a cubic coupling that takes the form of a Wigner 3-$j$ symbol \cite{Amit:1979ev,Benedetti:2020iku,Nador:2023inw}. However, now that we have specified the result of these melonic mechanisms, which is melonic dominance, we need not discuss them any further: see the various reviews \cite{Rosenhaus:2018dtp,Gurau:2019qag,Benedetti:2020seh} for details of their implementation.

The long-range models are not considered here, because they have the same scaling dimension as these short-range models -- by construction \cite{Gross:2017vhb,Benedetti:2019rja,Benedetti:2020rrq,Benedetti:2021wzt,Shen:2023srk}. Likewise, we will not be investigating the renormalization group (RG) flow between these CFTs, and the associated QFTs \cite{Berges:2023rqa, Benedetti:2019rja,Fraser-Taliente:2024rql}, as we work strictly in the conformal limit. %

\subsection{Conventions for representations}

We use the following conventions for the representations of the fields in our Euclidean QFTs. We denote a field in the $\SO(d)$ representation $\rho_\Phi$ by $\Phi_{\mu_1 \cdots \mu_s}$, where $\mu_i$ are the generalized $\SO(d)$ indices -- that is either vectorial or spinorial. This field may also transform in a representation $R_\Phi$ of some finite internal symmetry group $\Gglobal$ (i.e. $\dim \Gglobal$ does not scale with $N$), which we assume throughout to be unbroken.
Then we say that the field transforms in the representation
\begin{equation}\begin{aligned}
\rhoext_\Phi=\rho_\Phi \times R_\Phi \text{ of }\SO(d)\times \Gglobal.
\end{aligned}\end{equation} 
For example, for an $\gO(N)$ vector of complex Dirac fermions $\psi_i$ we have
\begin{equation}\begin{aligned}
\rhoext_{\psi} = (\text{Dirac fermion of }\SO(d), \text{vector of } \mathrm{O}(N))
\end{aligned}\end{equation} 
so that the real dimension of $\rhoext_\psi$ is
\begin{equation}\begin{aligned}
\dim \rhoext_{\psi} = \dim_{\mathbb{R}}\rho_\psi \times\dim R_\psi=  2 \Tr[\mathbb{I}_{s}] \times N.
\end{aligned}\end{equation}
The relevant data for a melonic CFT are the Lorentz representations $\rho_\Phi$ and the dimensions of the global symmetry representations $\dim R_\Phi$ -- other details of the $\Gglobal$ representations do not matter.

\section{A review of the \FtextOrPDF-theorems and \FttextOrPDF-maximization} \label{sec:FthmsFmax}

\subsection{The free energy in QFT and \texorpdfstring{$C$}{C}-functions} %

In the Wilsonian renormalisation group, each RG step is composed of a Kadanoff blocking followed by a rescaling.
The Kadanoff blocking clearly decreases the number of degrees of freedom towards the IR; however, the rescaling step reintroduces degrees of freedom, complicating this interpretation.
Nonetheless, we still need to capture our intuition that the number of effective degrees of freedom of a QFT decreases under RG flow.
That is, we want to define a $C$-function (in the sense of Cardy \cite{Cardy:1988cwa}) such that $\odv{C(E)}{E} > 0$.
This is hard, so a slightly weaker condition is to require a \textit{weak $C$-function}, defined only at the ends of the flow (where generically we expect a CFT), such that $C_{UV} > C_{IR}$.

In statistical mechanics at finite temperature, the partition function counts the number of active configurations -- that is, those with $E \lesssim T$:
\begin{equation}\begin{aligned}
Z = \sum_i e^{-E_i/T}.
\end{aligned}\end{equation} 
The free energy $F=-\log Z$ in QFT is then a monotonic reparametrisation of the number of active configurations. %
Unfortunately, $F$ is not a suitable weak $C$-function for Euclidean CFTs, as it is mired in subtleties, possessing a volume divergence $\int \dd^d x$, cutoff dependence, scheme dependence, and gauge dependence. All of these obscure the universal information hiding within it. $\Ft$, a modified version of the \textit{sphere} free energy, defined below, evades these issues -- thus providing a candidate weak $C$-function.

\subsection{\FttextOrPDF and the generalized \FttextOrPDF-theorem}

For a given CFT in continuous dimension $d$, we define 
\begin{equation}\begin{aligned}\label{eq:Ftdef}
\Ft \equiv -\sin(\pi d/2) F = \sin(\pi d /2) \log Z_{S^d},
\end{aligned}\end{equation}
where $Z_{S^d}$ is the partition function evaluated on the sphere $S^d$. The computation on the sphere regulates the volume divergence. The computation in continuous dimension removes the cutoff dependence of $F=-\log Z_{S^d}$, except when approaching even dimensions $d=2\mathbb{N}-\epsilon$, where the Weyl anomaly gives a $\sim a/\epsilon$ divergence in $F$. Essentially, this procedure involves analytically continuing in $d$ to dimension low enough that all power-law divergences in $F$ vanish. %
We notice that we can make this finite in all $d$ by dividing through by the volume of Euclidean hyperbolic space, $\vol H^{d+1}$. The dimensions are kept the same by actually multiplying by $\half \vol S^{d+1}/\vol H^{d+1}$: this yields the overall factor of $-\sin(\pi d/2)$ in \eqref{eq:Ftdef} \cite{Giombi:2014xxa}.%

$\Ft$ interpolates between $(-1)^{d/2} \pi a/ 2$ for the Weyl anomaly coefficients $a$ in even dimension, and $(-1)^{(d-1)/2} \log Z_{S^d}$ in odd dimensions. In dimensions $2,3$, and $4$ it is indeed a weak $C$-function, because so are the Weyl anomalies in $d=2,4$ and $F=-\log Z_{S^3}$ in $d=3$ \cite{Zamolodchikov:1986c,Komargodski:2011vj, Pufu:2016zxm}. %
The conditions for $\Ft$ to be a weak $C$-function in continuous dimension are not clear, and there are trivial counterexamples from the flow between generalized free field theories which violate the unitarity bound \cite{Benedetti:2021wzt}.
Nonetheless, in the following we shall show that $\Ft$ can be used in generic dimension to determine the IR limit of certain large-$N$ theories via an extremization principle. %

\subsection{Generalized free fields and \FttextOrPDF}

We briefly review generalized free fields (GFFs), following \cite{Benedetti:2021wzt}. These are essentially free fields with arbitrary propagators. If a theory containing GFFs is conformal, we call it a conformal generalized free field theory (GFFT), or a long-range massless Gaussian theory. A theory of GFFs is a mean field theory (MFT), as all correlators are simply sums of products of two-point functions. 
Note that it is a generic result that MFT is the leading contribution in large-$N$ theories, due to factorization \cite{Karateev:2018oml}. %
Once we have a particular mean field theory CFT, it is straightforward to conformally map it to the sphere; this enables our calculation of $\Ft$.

\subsubsection{Generalized free fields}

The standard free bosonic field, with a local action, has scaling dimension $\Delta=\frac{d-2}{2}$. By giving up locality, we can equally consider the case of a free field with arbitrary $\Delta$. Then the action in flat space is given by
\begin{equation}\begin{aligned}
S &=\half \int \dd^d x \, \phi(x) (-\partial^2)^{\frac{d}{2}-\Delta} \phi(x) = \half \int \frac{\dd^d p}{(2\pi)^d} \, \tilde{\phi}(p) (p^2)^{\frac{d}{2}-\Delta} \tilde{\phi}(-p).
\end{aligned}\end{equation} 
We will prefer to use the following bilocal form, which essentially contains the explicitly non-local implementation of the operator $(-\partial^2)^{\frac{d}{2}-\Delta}$,
\begin{equation}\begin{aligned}
S &= \lim_{r\to 0} \half \int_{\abs{x-y}>r} \dd^d x \, \dd^d y\, \phi(x) G_\phi^{-1}(x,y) \phi(y),\\
\end{aligned}\end{equation} 
though we will hide the limit in all subsequent computations. The inverse propagator is
\begin{equation}\begin{aligned}
G_\phi^{-1}(x,y) \equiv \frac{c(d-\Delta)}{\abs{x-y}^{2(d-\Delta)}}, \quad c(\Delta) \equiv \frac{1}{2^{d-2\Delta} \pi^{d/2}} \frac{\Gamma(\Delta)}{\Gamma(\frac{d}{2}-\Delta)},
\end{aligned}\end{equation} 
which manifestly gives a conformal correlator. The propagator is then
\begin{equation}\begin{aligned} \label{eq:flatSpaceGFFpropagator}
G_\phi(x,y) = \frac{c(\Delta)}{\abs{x-y}^{2\Delta}},
\end{aligned}\end{equation} 
which can be shown by Fourier transforming twice. Due to the quadratic form of the action, all higher-point correlators of a theory consisting solely of GFFs can be found by Wick contractions, in the usual manner for a free theory.

We now conformally map \eqref{eq:flatSpaceGFFpropagator} to the sphere of radius $R$, via the stereographic projection.  %
Recall that the $d$-dimensional sphere $S^d$ with a single point removed is conformally equivalent to Euclidean space: the conformal map is the stereographic projection. %
Hence, the sphere is conformally flat. %
This makes the map of the flat-space CFT to the sphere CFT straightforward (up to subtleties associated with the curvature couplings, which we can neglect for GFFs \cite{Pufu:2016zxm}). %
In stereographic coordinates, the sphere metric is 
\begin{equation}\begin{aligned} \label{eq:sphereMetric}
g_{\mu\nu} = \Omega(x)^2 \delta_{\mu\nu}, \quad \quad \Omega(x) \equiv \frac{2 R}{(1+x^2)^\half},
\end{aligned}\end{equation} 
and the chordal distance $s(x,y)$ is
\begin{equation}\begin{aligned}
s(x,y) \equiv \Omega(x)^\half\Omega(y)^\half \abs{x-y}.
\end{aligned}\end{equation} 
To conformally map the propagator, we simply replace $\abs{x-y} \to s(x,y)$ in \eqref{eq:flatSpaceGFFpropagator} to get
\begin{equation}\begin{aligned} \label{eq:sphereGFFpropagator}
G_\phi(x,y)|_{S^d} = \frac{c(\Delta)}{s(x,y)^{2\Delta}}.
\end{aligned}\end{equation} 
The same applies for fields in other Lorentz representations. However, we note that to reproduce \eqref{eq:sphereGFFpropagator}, the sphere action will generically need additional non-minimal couplings to the geometry\footnote{For example, in the case of the conformally coupled scalar of dimension $\Delta= \frac{d-2}{2}$, the sphere action is $$S_{\phi}=\int_{S^d} \dd^d x \sqrt{g} \, \half \phi \left(-\partial^2 + \frac{d-2}{4(d-1)} R\right) \phi.$$}.

\subsubsection{Free energy}
\begin{figure}
\centering
    \begin{subfigure}{0.5\textwidth}
\includegraphics[width=\textwidth]{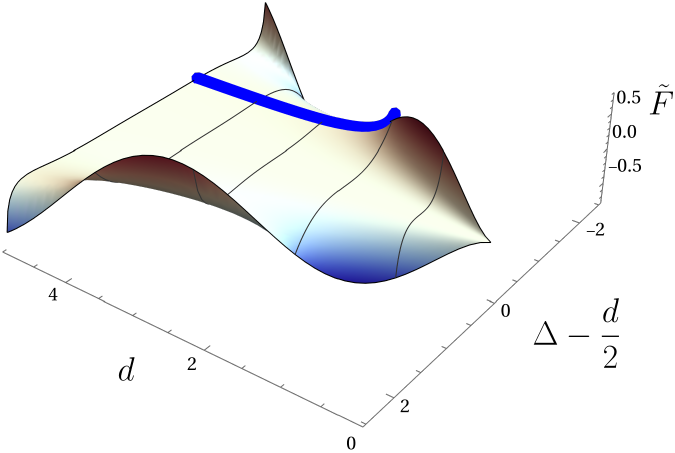}
    \caption{$\Ft$ for a free boson}
    \end{subfigure}%
    \begin{subfigure}{0.5\textwidth} 
\includegraphics[width=\textwidth]{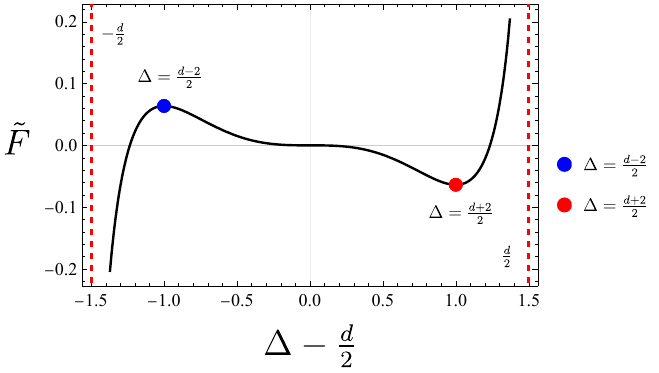}
    \caption{Cross-section for $d=3$}
    \label{fig:Ftdeq3}
    \end{subfigure}
\caption{$\Ft_b(\Delta)$ for a generalized free scalar field, \eqref{eq:Ftboson}, shown as a surface for $0+\epsilon<\Delta<d-\epsilon$. The function is even around $\Delta=\frac{d}{2}$. The blue line indicates the free field solution $\Delta = \frac{d-2}{2}$. $\Ft(\Delta)$ looks schematically like \cref{fig:Ftdeq3} ($d=3$), for all values of $d$: there is always a local maximum at the scaling dimension of the free field $\frac{d-2}{2}$; a stationary point at $\frac{d}{2}$; and a local minimum at the shadow of the free field $\frac{d+2}{2}$. There is also always a log-divergence for $\Delta$ approaching $0,d$.}  %
\label{fig:Ftboson3D}
\end{figure}

The free energy of a generalized free bosonic field on the sphere is
\begin{equation}\begin{aligned}
F_\phi &= -\log Z_{\phi,S^d}= \log \int \cD \phi\, \exp(- \int_{x,y}\half \phi(x) G_\phi^{-1}(x,y) \phi(y)) \\
&= \half\log\det G_\phi^{-1} = \half \Tr\log G_\phi^{-1},
\end{aligned}\end{equation} 
where we have defined the path-integral measure to cancel any constant factors, and $G_\phi^{-1}$ is the inverse sphere propagator. For a field of arbitrary statistics, we need only modify this to $F_\phi=(-1)^{\mathrm{F}_\phi} \half \Tr\log G_\phi^{-1} \equiv \half \mathrm{Str}\log G_\phi^{-1}$ (taking $\mathrm{F}_\phi=0$ for bosons and $1$ for fermions). Hence,
\begin{subequations} \label{eq:FtBosFerm}
\begin{equation}\begin{aligned}
\Ft_{\phi} = -\sin(\pi d/2) \, \half \,\mathrm{Str} \log G_{\phi}^{-1}(x,y)
\end{aligned}\end{equation} 
for a real field with sphere propagator $G_{\phi}(x,y)$. 

The computation of $\Ft_\phi$ when $G_\phi$ is the conformal sphere propagator is a standard result from the AdS/CFT literature for any Lorentz representation of $\phi$ \cite{Sun:2020ame,Benedetti:2021wzt,Harribey:2022esw}. %
 We simply quote the results, which can be found straightforwardly for arbitrary $d$ using zeta function regularisation. %
For a free real scalar boson of dimension $\Delta$, we have
 \begin{equation}\begin{aligned} \label{eq:Ftboson}
\Ft_b(\Delta, \text{scalar}) \equiv \frac{\pi}{\Gamma(d+1)}\int_{\frac{d}{2}}^{\Delta} \dd \Delta' \, \frac{\Gamma (\Delta' ) \Gamma (d-\Delta' )}{\Gamma \left(\frac{d}{2}-\Delta' \right) \Gamma \left(\Delta' -\frac{d}{2}\right)},
 \end{aligned}\end{equation} 
 where we have used $\Ft(\Delta=d/2)=0$. The behaviour of this function in continuous dimension for $0<\Delta<d$ is shown in \cref{fig:Ftboson3D}. %
For a complex fermionic field of dimension $\Delta$, %
\begin{equation}\begin{aligned}
\Ft_f(\Delta, \text{Dirac f.}) &\equiv %
-2\Tr\spinid \frac{\pi}{\Gamma (d+1)}\int_{\frac{d}{2}}^{\Delta} \dd \Delta' \frac{\Gamma \left(\Delta' +\frac{1}{2}\right) \Gamma \left(d-\Delta' +\frac{1}{2}\right)}{\Gamma \left(\frac{d}{2}-\Delta' +\frac{1}{2}\right) \Gamma \left(-\frac{d}{2}+\Delta' +\frac{1}{2}\right)},
\end{aligned}\end{equation} 
\end{subequations}
where we have used
\begin{equation*}
\dim\rhoext_\psi = \dim \rho_\psi \times \dim R_\psi = \dim(\text{Dirac fermion}) = 2 \Tr\spinid.
\end{equation*}
The trace structure of $\Ft_\phi$ makes it easy to see the following two facts. First, the results \eqref{eq:FtBosFerm} should never depend on the normalization of the field $\phi$ -- indeed they do not, because $\Tr\log(C\delta^d(x-y))= 0$, for any constant $C$. Second, it is trivial that $\Ft_\phi$ for a field $\phi$ in a representation $\rhoext=\rho \times R$ of $\SO(d) \times \Gglobal$ is %
\begin{equation}\begin{aligned}
\Ft(\Delta, \rhoext) = \dim R \times \Ft(\Delta, \rho).
\end{aligned}\end{equation} 
For example, a complex scalar has $\Ft(\Delta, \text{complex scalar}) = 2 \Ft(\Delta, \text{scalar})$.
Finally, as demonstrated in \cref{fig:Ftboson3D}, for all $d$, $\Ft_b(\Delta)$ for the scalar GFF has the following properties. $\Ft$ hits a maximum for the free field, $\Delta=\frac{d-2}{2}$, with a corresponding minimum for the shadow of the free field at $\frac{d+2}{2}$ -- and a stationary point in between at $\frac{d}{2}$. The same applies for the fermion, with free field $\Delta = \frac{d-1}{2}$.
Hence, the free field values are always local maxima, and certainly absolute maxima within the range given by the free field value and its shadow. %

\subsection{\FttextOrPDF-maximization in superconformal field theories} \label{sec:Fmaximization}%

We now review the $F$- and $a$-maximization procedures in SCFTs; as mentioned in the introduction, we unify these procedures by considering $\Ft$-maximization in general dimension \cite{Giombi:2014xxa, Minahan:2015any}.

Take a UV supersymmetric QFT in $d$-dimensions with four supercharges, containing free chiral superfields $\{X\}$, in some general global symmetry representations. Then perturb by some superpotential $W(\{X\})$ that preserves the supercharges and global symmetry.  %

Now flow the theory to the deep IR; assuming that there is an SCFT there, given certain caveats (see \cref{sec:FmaxLimitations}), the $R$-symmetry generator in the IR is a linear combination of the original $UV$ generator and any other abelian symmetry generators. The supersymmetry then enforces the relationship 
\begin{equation}\begin{aligned} \label{eq:DeltaRrel}
\Delta_X = \frac{d-1}{2} R_X
\end{aligned}\end{equation} 
between the scaling dimensions and the $R$-charges. We can find the IR scaling dimensions by extremizing the total $\Ft$ for a collection of generalized free chiral superfields with \textit{trial} scaling dimensions $\Delta_{X}$
\begin{equation}\begin{aligned}
\Ft = \sum_{\text{chirals } X} \tilde{\mathcal{F}}(\Delta_{X}), \quad \tilde{\mathcal{F}}(\Delta) \equiv \Ft(\Delta, \text{chiral superfield}), \label{eq:chiralFtSumRepeated}
\end{aligned}\end{equation} 
subject to a constraint. This constraint is implied by insisting that we have supersymmetry and an $R$-charge in the IR, which fixes the $R$-charge of the superpotential to be two. %
This constraint can be rewritten suggestively in terms of the scaling dimensions $\Delta_{X}$ using the relationship \cref{eq:DeltaRrel}: assuming a perturbing superpotential of the form
\begin{equation}\begin{aligned}
 S_{\text{int}} \supset \int \dd^d x \dd^2 \theta \, W, \quad W \equiv \sum_m g_m\prod_{\text{chirals }X} X^{q_{X}^m}, %
\end{aligned}\end{equation} 
then, for each monomial, we have the constraint
\begin{equation}
\sum_{\text{chirals } X} q^m_{\phi} \Delta_{X} + [\dd^d x \dd^2 \theta] =0, \quad [\dd^d x \dd^2 \theta] = -(d-1). \label{eq:chiralsConstraintsRepeated}
\end{equation}

Each chiral superfield contains a complex scalar, a complex auxiliary field, and a complex Dirac fermion; so $\tilde{\cF}(\Delta_X)$ is defined by summing up the contributions of each of these components,
\begin{equation}\begin{aligned}
\Ft(\Delta) &= \Ft_b(\Delta, \text{complex scalar}) + \Ft_f(\Delta+\thalf, \text{Dirac fermion}) +  \Ft_b(\Delta+1, \text{complex scalar}),
\end{aligned}\end{equation} 
up to a convention-dependent additive constant which drops out of all computations. To ensure supersymmetry, we take $\Tr\spinid=2$ fixed, regardless of dimension: this is the dimensional reduction scheme, a standard procedure for analytically continuing $3$d supersymmetry away from $d=3$ \cite{Giombi:2014xxa}. %

These constraints \eqref{eq:chiralsConstraintsRepeated} correspond exactly to the requirement that the superpotential is marginal in the IR. This is because the scaling dimensions of products of chiral superfields add linearly (precisely due to the relationship \cref{eq:DeltaRrel}), and the form of the superpotential is protected by supersymmetry: hence the scaling dimension of each monomial is indeed $\sum_\phi q^m_\phi \Delta_\phi$. Thus, as in the melonic case, we extremize $\Ft$ for a mean field theory, subject only to the IR marginality of the potential \eqref{eq:chiralsConstraintsRepeated}. If the SCFT is also unitary, it turns out that the extremum is also a maximum; whether this holds in any sense for the melonic theories as well is not yet clear. %

\subsubsection{Simple example}

Consider the example of $N+1$ chiral superfields, and a potential $W =\frac{\lambda}{2} X \sum_{i=1}^N Z_i Z_i$. To find the exact IR scaling dimensions, we maximise
\begin{equation}\begin{aligned}
\Ft = \tilde{\cF}(\Delta_X) + N \tilde{\cF}(\Delta_Z), \text{ subject to }\Delta_X + 2\Delta_Z = d-1,
\end{aligned}\end{equation} 
with \text{no} $N$-subleading corrections!

\subsubsection{Limitations of \FttextOrPDF-maximization} \label{sec:FmaxLimitations}

The principal limitation with $\Ft$-maximization is we must assume that the flow ends at an SCFT without any accidental symmetries arising in the IR. If this is not the case, the $R$-charge can mix with said symmetries. We can see this in a slight modification of the example above, and consider the UV perturbation
\begin{equation}\begin{aligned}
W = g_1 X \sum_{i=1}^N Z_i Z_i + g_2 X^3.
\end{aligned}\end{equation} 
This triggers an RG flow, which preserves a $\gO(N) \times \mathbb{Z}_3$ flavour symmetry and a unique $\gU(1)_R$ symmetry, under which all the fields have $R$-charge $2/3$. However, for $N>2$, at least, the coupling $g_2$ is thought to run to zero; the flavour symmetry is then enhanced to $\gO(N) \times \gU(1)$, and so the IR $R$-charges, and so scaling dimensions, are not determined by the naive constraint $3 \Delta_X = d-1$ that would come from a non-zero $g_2$ \cite{Pufu:2016zxm}.

This phenomenon has a parallel in the case of the melonic-type theories: essentially, it corresponds to the fact that when we have multiple melonic interactions $g_m$, any of them can either be tuned to zero, or could run to zero. %

\section{Fundamental claim} \label{sec:fundamentalClaim}

\begin{mdframed}
\textbf{The complicated IR structure of melonic field theories, with arbitrary number of fields and interactions, can be reformulated as a constrained $\tilde{F}$-extremization problem.} The defining data for these melonic theories is
\begin{enumerate}
    \item a list of $n_f$ fields and representations (Lorentz representations, and global symmetry representations -- $(\phi, \rho_{\phi}, R_{\phi})$);
    \item an $n_m \times n_f$ matrix of integers -- $q^m_{\phi}$, determined from the schematic form of the melonic-dominant potential $V = \sum_{m=1}^{n_m} g_m \prod_{\phi} \phi^{q^m_\phi}$. 
\end{enumerate}
Then, we can compute the IR scaling dimensions by extremizing the free energy $\Ft=-\sin(\pi d/2)F$ of the mean field theory consisting of a collection of generalized free fields
\begin{subequations} \label{eq:FmaxSummary}
\begin{equation}
\Ft(\{\Delta_{\phi}\}) = \sum_{\text{fields } \phi} \Ft_{\phi}(\Delta_{\phi}, \rhoext_{\phi}) = \sum_{\text{fields } \phi} \Ft_{\phi}(\Delta_{\phi}, \rho_{\phi}) \times \dim R_{\phi},
\end{equation} 
with respect to the trial scaling dimensions $\Delta_{\phi}$, subject to the melonic constraints
\begin{equation} \label{eq:melConstrsIn4}
\sum_{\text{fields } \phi} q^m_{\phi} \Delta_{\phi} -d =0
\end{equation} 
for each of the couplings $g_m$. Essentially, these constraints enforce the marginality of $V$ in the IR. This extremization will typically give a discrete infinity of solutions -- consistency with the UV description then requires that the scaling dimensions of the fields must be greater than their free (UV) scaling dimensions.
Thus,
\begin{equation} \label{eq:DelGreaterThanFree}
\Delta_{\phi} > \Delta^\mathrm{free}_{\phi},\\
\end{equation} 
\end{subequations}
where for dynamical scalars $\Delta_\phi^\mathrm{free} = \frac{d-2}{2}$; for fermions $\Delta_\psi^\mathrm{free} = \frac{d-1}{2}$; and for auxiliary scalars $\Delta_X^\mathrm{free} = \frac{d}{2}$.
Otherwise, the free behaviour will dominate in the IR; of course, for the interacting fields \eqref{eq:DelGreaterThanFree} also corresponds to the unitarity bound in $d \ge 2$.
Therefore, only $\{\Delta_{\phi}\}$ lying inside a polyhedron in $\mathbb{R}^{n_f}$, which we call the \textit{IR wedge}, are valid solutions\footnote{Solutions satisfying $\Delta_{\phi} < \Delta^\mathrm{free}_{\phi}$  -- lying in the \textit{UV wedge} -- are also consistent as UV CFTs; however, we will usually neglect them.}.

Note that it is possible for the constraint system to be over-determined (if $n_m > n_f$). In that case, to find an IR solution, some of the couplings $g_m$ must either be set to zero, or run to zero. In the latter case, IR consistency demands that $\sum_{\phi} q^m_{\phi} \Delta_{\phi} > d$. We will typically assume that we keep only the $g_m$s that are non-zero at the fixed point.
\end{mdframed}
Other than the substitution of $[\dd^dx \dd^2\theta]=d-1$ with $[\dd^dx]=d$, this is precisely identical to the supersymmetric $\Ft$-maximization of \cref{sec:Fmaximization}, except that, as far as we know, the solution need not only be a maximum. %

Incorporating the constraints \eqref{eq:melConstrsIn4} into $\Ft$, using Lagrange multipliers $\mathfrak{g}_m'$, we find
\begin{equation}\begin{aligned}
\frac{1}{N}\Ft(\{\Delta_{\phi}, \mathfrak{g}_m'\}) = \sum_{\text{fields } \phi} \Ft_{\phi}(\Delta_{\phi}) +\sum_{\text{melons } m} \mathfrak{g}_m' \left(\sum_{\text{fields } \phi} q^m_{\phi} \Delta_{\phi} -d\right).
\end{aligned}\end{equation} 
We have suppressed the dependence on the representations. We will show in the next section that this is an expansion of the sphere free energy of the melonic CFT, where $\mathfrak{g}_m'$ is proportional to the running coupling constant squared (that is, including the field renormalizations).
Extremizing this quantity with respect to the $\Delta_{\phi}$s and $\mathfrak{g}_m'$s then determines the conformal scaling dimensions of the theory. The Lagrange multipliers enforcing marginality can therefore be given a precise interpretation as being the squared running coupling constants; this is precisely the conjecture of Kutsakov \cite{Kutasov:2003ux,Barnes:2004jj,Amariti:2011xp} in the case of $a/F$-maximization in SCFTs.

The final results are a function only of the discrete data of the integers $q^m_{\phi}$ and the representations of the fields. In generic dimension, this $\Ft$-extremization procedure must be done numerically, typically yielding multiple possible IR vacua: we now demonstrate this with an explicit example.

\subsection{Explicit example: the melonic quartic Yukawa model}

\begin{figure}
    \centering
    \includegraphics[width=0.7\textwidth]{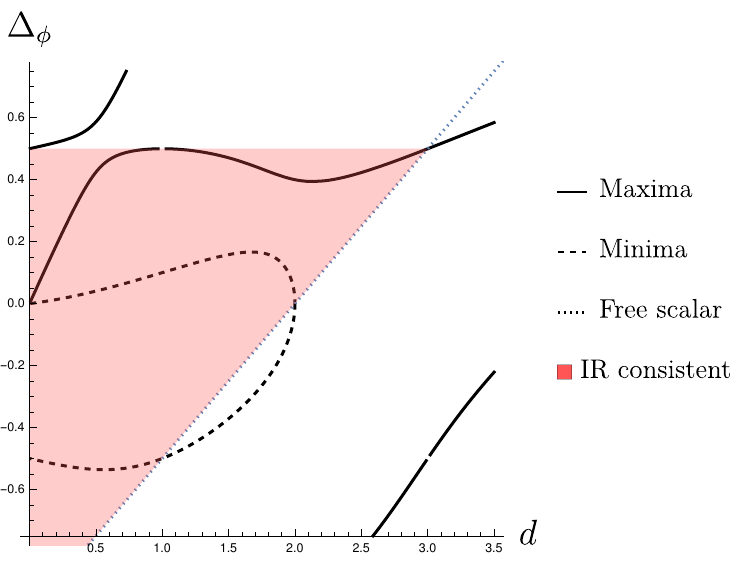}
    \caption{The scaling dimension of $\phi$ for the melonic quartic Yukawa theory \eqref{eq:lmelonicLagrangian} in continuous dimension; this follows from extremizing $\Ft(\Delta_\phi, \Delta_\psi, \mathfrak{g}')$ in \eqref{eq:lmelonicFt}. $\Delta_\psi$ can be found by the constraint $2\Delta_\phi + 2\Delta_\psi = d$. The red region indicates the IR-consistent wedge $\Delta_\phi \ge \frac{d-2}{2}$ and $\Delta_\psi \ge \frac{d-1}{2}$; this theory is free in the upper corner of this wedge, for $d=3$. Solid lines are maxima of the constrained $\Ft(\Delta_\phi) =\Ft_b(\Delta_\phi)+ \Ft_f(\tfrac{d-2\Delta_\phi}{2})$; dashed lines are minima. }
    \label{fig:phi2psibarpsiIRwedge}
\end{figure}

Consider the melonic quartic Yukawa model (studied in detail in \cite{Fraser-Taliente:2024rql}): it is the theory of $N$ Dirac fermions\footnote{As in the supersymmetric case, we take the gamma matrices to have dimension $2$, regardless of $d$.} and $N$ bosons that is marginal in $d=3$, with schematic Lagrangian
\begin{equation}\begin{aligned} \label{eq:lmelonicLagrangian}
\cL_{QY} = \half \phi(-\partial^2)\phi +\bar\psi (-\slashed{\partial}) \psi + \half g\phi\phi\bar\psi\psi.
\end{aligned}\end{equation} 
Note that we have suppressed the particular melonic mechanism here, which could be a disorder average, as in SYK \cite{Prakash:2022gvb}, or tensorial \cite{Fraser-Taliente:2024rql}. The scalar potential has also been tuned to zero. Therefore, we extremize 
\begin{equation}\begin{aligned} \label{eq:lmelonicFt}
\Ft(\Delta_\phi, \Delta_\psi, \mathfrak{g}')/N = \Ft(\Delta_\phi, \text{scalar}) + \Ft(\Delta_\psi, \text{Dirac fermion}) -\frac{\mathfrak{g}'}{2} (2\Delta_\phi + 2\Delta_\psi -d)
\end{aligned}\end{equation} 
with respect to the $\Delta$s and the Lagrange multiplier $\mathfrak{g}'$, while also requiring $\Delta_\phi > \frac{d-2}{2}$ and $\Delta_\psi > \frac{d-1}{2}$.  Conveniently, we are taking derivatives with respect to $\Delta$, so do not need to do the integrals in \eqref{eq:FtBosFerm}.
After a numerical extremization procedure we obtain the results given in \cref{fig:phi2psibarpsiIRwedge}, where the IR wedge (evidently bounded by the scaling dimension of a free scalar) is shown in red, $\frac{d-2}{2} < \Delta_\phi < \half$. Since we are extremizing with respect to only a single variable $\Delta_\phi$, we only have a single constraint; thus all extrema are either maxima or minima. The line descending here from the free theory ($\Delta_\phi = \frac{d-2}{2}, \Delta_\psi = \frac{d-1}{2}$) that exists in $d=3$ is then indeed a maximum; however, for $d \le 2$, we also have two lines of minima. We comment on this model further in \cref{sec:QuarticYukawaFull}.

\section{\FttextOrPDF-extremization from the 2PI effective action} \label{sec:FtMaxFrom2PI}

This mechanism is a straightforward consequence of the 2PI formalism, which we first briefly review. We aim to reach a simple expression \eqref{eq:Fin2PI} for the sphere free energy $F$ as a function of $G=\expval{\varphi(x) \varphi(y)}^\mathrm{conn}$, the full propagator. For more details, see \cite{Benedetti:2018goh} and references therein.

\subsection{Reminder of the 2PI formalism}

The two-particle irreducible (2PI) effective action $\Gamma[\phi,G]$ is defined as the double Legendre transform of the generating functional $\mathbf{W}[j,k]$ with respect to $j$, a one-point source, and $k$, a two-point source. Explicitly, for the theory of a scalar field with action $S[\varphi]$,
\begin{equation}\begin{aligned}
\mathbf{W}[j,k] \equiv \log \int \cD \varphi \, \exp\left(- S[\varphi] + \int_x j(x) \varphi(x) + \int_{x,y} \half \varphi(x) k(x,y) \varphi(y)\right).
\end{aligned}\end{equation} 
This is precisely (minus) the sourced free energy. Now, we define the expectation value of the field $\varphi$ in the presence of the sources
\newcommand{\bfPhi}{\Phi}
\begin{equation}\begin{aligned} \label{eq:classicalFieldDef}
\bfPhi(x) = \expval{\varphi(x)}_{j,k} = \fdv{\mathbf{W}}{j(x)};
\end{aligned}\end{equation} 
and the full connected propagator%
\begin{equation}\begin{aligned} \label{eq:GDef}
\mathbf{G}(x, y) &= \expval{\varphi(x) \varphi(y)}^{\mathrm{conn}}_{j,k} = \expval{\varphi(x)\varphi(y)}_{j,k} -\expval{\varphi(x)}_{j,k}\expval{\varphi(y)}_{j,k}\\
& = 2 \fdv{\mathbf{W}}{k(x,y)} - \fdv{\mathbf{W}}{j(x)}\fdv{\mathbf{W}}{j(y)},
\end{aligned}\end{equation} 
also in the presence of the sources. The (double) Legendre transform with respect to both $j$ and $k$ is implemented almost as usual \cite{Zia:2008LegendreTransform}, giving the 2PI effective action\footnote{We overload notation and use the same symbol to refer to a given object, regardless of whether it is the argument of a functional or a functional itself. Hence, we have either $(j,k,\bfPhi[j,k],\mathbf{G}[j,k])$ or $(j[\bfPhi, \mathbf{G}],k[\bfPhi, \mathbf{G}], \bfPhi, \mathbf{G})$; regardless of which we take, \eqref{eq:GammaAndWLT} holds.}: %
\begin{equation}\begin{aligned} \label{eq:GammaAndWLT}
\Gamma[\bfPhi, \mathbf{G}] + \mathbf{W}[j,k] = \int_x j(x) \bfPhi(x) + \half \int_{x, y} (\mathbf{G}(x, y) + \bfPhi(x) \bfPhi(y)) k(x, y).
\end{aligned}\end{equation} 
Hence, in the unsourced case $(j=0,k=0)$, $\Gamma$ coincides precisely with the free energy. The reason for the additional $\half$ and $\bfPhi(x) \bfPhi(y)$ in the second Legendre transform is to ensure that we get the connected $\mathbf{G}$ defined in \cref{eq:GDef}, rather than the disconnected propagator.

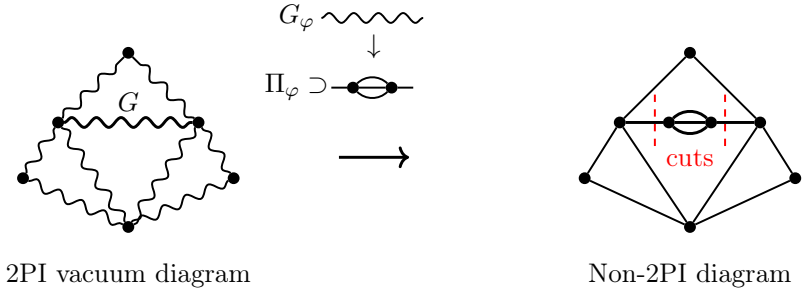
\begin{figure}
    \centering
    \begin{tikzpicture}[
    thick,
    vertex/.style={draw, circle, fill=black, inner sep=0pt, minimum size=4pt},
    Gprop/.style={decorate, decoration={snake, amplitude=2pt}, draw=black, thick},
    regular/.style={draw=black},
    label distance=0.5cm
]
\begin{scope}[xshift=0cm]
    \node[vertex] (v1) at (0,0) {};
    \node[vertex] (v2) at (2,0) {};
    \node[vertex] (v3) at (1,1) {};
    \node[vertex] (v4) at (1,-1.5) {};
    \node[vertex] (v5) at (-0.5,-0.8) {};
    \node[vertex] (v6) at (2.5,-0.8) {};
    
    \draw[Gprop,very thick] (v1) -- (v2);
    \draw[Gprop] (v1) -- (v3);
    \draw[Gprop] (v2) -- (v3);
    \draw[Gprop] (v1) -- (v4);
    \draw[Gprop] (v2) -- (v4);
    \draw[Gprop] (v5) -- (v1);
    \draw[Gprop] (v6) -- (v2);
    \draw[Gprop] (v5) -- (v4);
    \draw[Gprop] (v6) -- (v4);
    
    \node at (1,-2.2) {2PI vacuum diagram};
    \node at (1,0.3) {$G$};
\end{scope}

\draw[->, very thick] (4,-0.5) -- (5,-0.5);
\begin{scope}[yshift=1.5cm]
    \draw[Gprop] (3.75,0) -- (5.2,0); %
    \node at (3.4,0) {$G_\varphi$};
    \node at (3.4,-0.99) {$\Pi_\varphi \supset $}; %
    \node at (4.5,-0.4) {$\downarrow$};
    \node (b1) at (3.75,-1) {};
    \node (b2) at (5.2,-1) {};
    \node[vertex] (s1) at (4.2,-1) {};
    \node[vertex] (s2) at (4.75,-1) {};
    \draw[regular] (b1) -- (s1) -- (s2);
    \draw[regular] (s1) to[bend left=45] (s2);
    \draw[regular] (s1) to[bend right=45] (s2);
    \draw[regular] (s2) -- (b2);
\end{scope}

\begin{scope}[xshift=8cm]
    \node[vertex] (w1) at (0,0) {};
    \node[vertex] (w2) at (2,0) {};
    \node[vertex] (w3) at (1,1) {};
    \node[vertex] (w4) at (1,-1.5) {};
    \node[vertex] (w5) at (-0.5,-0.8) {};
    \node[vertex] (w6) at (2.5,-0.8) {};
    
    \node[vertex] (s1) at (0.7,0) {};
    \node[vertex] (s2) at (1.3,0) {};
    
    \draw[regular, very thick] (w1) -- (s1);
    \draw[regular, very thick] (s1) to[bend left=45] (s2);
    \draw[regular, very thick] (s1) -- (s2);
    \draw[regular, very thick] (s1) to[bend right=45] (s2);
    \draw[regular, very thick] (s2) -- (w2);
    
    \draw[regular] (w1) -- (w3);
    \draw[regular] (w2) -- (w3);
    \draw[regular] (w1) -- (w4);
    \draw[regular] (w2) -- (w4);
    \draw[regular] (w5) -- (w1);
    \draw[regular] (w6) -- (w2);
    \draw[regular] (w5) -- (w4);
    \draw[regular] (w6) -- (w4);
    
    \draw[red, dashed, thick] (0.5,0.4) -- (0.5,-0.4);
    \draw[red, dashed, thick] (1.5,0.4) -- (1.5,-0.4);
    
    \node at (1,-2.2) {Non-2PI diagram};
    \node[red] at (1,-0.5) {cuts};
\end{scope}
\end{tikzpicture}
    \caption{Replacing a propagator for $\varphi$ in a 2PI diagram by a contribution to the field self-energy $\Pi_\varphi$ yields a diagram which automatically disconnects on cutting two lines. Hence, all such diagrams are not 2PI.}
    \label{fig:2PIdiagrams}
\end{figure}

Solving the theory corresponds to finding $\expval{\varphi}$ and $\expval{\varphi(x) \varphi(y)}^\mathrm{conn}$ in the unsourced case: we call these the classical field $\phi$ and the full two-point function $G$ respectively,
\begin{equation}\begin{aligned}
\phi \equiv \bfPhi|_{j=0,k=0}, \quad G \equiv \mathbf{G}|_{j=0,k=0}.
\end{aligned}\end{equation} 
The Legendre transform relations are that, considered as functionals of $\bfPhi(x)$ and $\mathbf{G}(x,y)$, $j[\bfPhi, \mathbf{G}]$ and $k[\bfPhi, \mathbf{G}]$ solve
\begin{subequations}\label{eq:LegendreTransformRelations}
\begin{align}\label{eq:EoMsForVEVs}
\fdv{\Gamma[\bfPhi , \mathbf{G}]}{\bfPhi(x)} &= j(x) + \int_y k(x,y) \bfPhi(y) , \\
\fdv{\Gamma[\bfPhi, \mathbf{G}]}{\mathbf{G}(x,y)} &= \half k(x,y).
\end{align}
\end{subequations} 
Therefore, the equations of motion for $\phi$ and $G$ are
\begin{equation}\begin{aligned}
j[\phi, G]=0, \,  k[\phi, G]=0 \, \implies \, \fdv{\Gamma[\phi, G]}{\phi}  = 0, \, \fdv{\Gamma[\phi, G]}{G} = 0.
\label{eq:varGeq0}
\end{aligned}\end{equation} 
In the melonic case, $\fdv{\Gamma}{\phi}=0$ is trivially solved by assuming no symmetry breaking, $\expval{\varphi}=0$; $\fdv{\Gamma}{G}=0$ will give the Schwinger-Dyson equation for the bilocal field $G$, which is the usual route to solve a melonic field theory. It is then a standard result that the 2PI effective action $\Gamma[\phi, G]$ for a scalar field theory is given by
\begin{subequations}\label{eq:2PIeffectiveActionScalarBoth}
\begin{equation}\label{eq:2PIeffectiveActionScalar}
\Gamma[\phi, G] = S[\phi] + \frac{1}{2} \Tr \ln G^{-1} + \half \Tr C^{-1} G  + \Gamma_{2}[\phi, G].
\end{equation}
Here:
\begin{itemize}
    \item $S[\phi]$ is the classical action, evaluated for the classical field $\phi$.
    \item $C^{-1}(t,t')$ is the (matrix) inverse free propagator for $\varphi$.
    \item As usual, we treat $G(x,y)$ and $C(x,y)$ as matrices indexed by $x$ and $y$, and take the matrix logarithm.
    \item $\Gamma_{2}[\phi, G]$ is (minus) the sum of all of the two-particle irreducible vacuum graphs. These are all graphs that do not disconnect when cutting open any two edges. Crucially, the Feynman rules are slightly modified: instead of the free $\varphi$ propagator, we use $G$. In the symmetric case where $\phi=0$, the vertices used are precisely the same as those in the original action (this is not the case if $\phi\neq 0$, i.e. if the field gets a VEV).  
\end{itemize}
This means that we never need to consider self-energy insertions: they are resummed automatically by the fact that we are using $G$! To see this, take any 2PI diagram, and replace one of the $\varphi$ propagators by a diagram that would contribute to the self-energy $\Pi_\varphi$ of that field. This is demonstrated in \cref{fig:2PIdiagrams}; clearly, in that case we can cut the two edges surrounding the $\Pi_\phi$ insertion and disconnect the diagram; thus it does not contribute to $\Gamma_{2}$.

The following schematic expression \cite{Benedetti:2018goh} is a useful aide-m\'emoire. If the original action was $S[\varphi]$, then we can write %
\begin{equation}\begin{aligned} \label{eq:integralFormForGamma}
e^{-\Gamma[\phi,G]} = e^{-S[\phi] - \half \Tr[C^{-1} G]} \int_{\mathrm{2PI}} \cD\varphi\, e^{-\half \varphi G^{-1} \varphi - S_{\mathrm{int}}[\phi, \varphi]},
\end{aligned}\end{equation} 
\end{subequations}
where the subscript indicates that when we do the perturbative expansion of the functional integral, we keep only the 2PI graphs. Here, $S_\mathrm{int}[\phi,\varphi]$ is the interacting part of $S[\phi + \varphi]$.

This formalism trivially generalizes to the multi-field case, with $\phi_i = \expval{\varphi_i}$ and $G_i = \expval{\varphi_i \varphi_i}^{\text{conn}}$; additionally, it is purely combinatoric, and therefore must also apply to a QFT on a sphere. Thus, we have all the ingredients we need to apply it to calculating the sphere free energy of any QFT, which we note from \eqref{eq:GammaAndWLT} is just
\begin{subequations}\label{eq:FeqGextremum}
\begin{equation}\begin{aligned} \label{eq:Fin2PI}
F \equiv -\log Z_{S^d} = \Gamma[\phi_i, G_i]|_{k_i=0, j_i=0, \text{ on }S^d},
\end{aligned}\end{equation} 
where $k=0$, $j=0$ just means that we evaluate $\Gamma$ for the $\phi_i$ and $G_i$s that solve \cref{eq:EoMsForVEVs},
\begin{equation}\begin{aligned}
\forall i \quad \fdv{\Gamma}{\phi_i} = 0, \quad \fdv{\Gamma}{G_i} = 0.
\end{aligned}\end{equation} 
\end{subequations}
\textbf{That is, the (sphere) free energy of any QFT is precisely the extremum of the 2PI action $\Gamma$, with respect to the propagators $G_i$ and field VEVs $\phi_i$.}

\subsection{Application of the 2PI formalism}

In the following, we calculate the sphere free energy in the IR, which is assumed to be conformal. Thus, we evaluate \eqref{eq:Fin2PI} for large radius $R$, such that the contribution of the UV propagator is negligible. We also assume no symmetry breaking of any kind, and so drop $\phi =\expval{\varphi}$ and $S[\phi]$ in \cref{eq:2PIeffectiveActionScalarBoth,eq:FeqGextremum}. In this case, we now find that for a melonic theory, \textbf{only $n_m$ diagrams appear in $\Gamma_2[G]$}: these are the complete melons
\begin{equation}
\Gamma_{2}[\{G_{\phi_i}\}] = -\half\sum_m \frac{g_m^2}{S_m} \int_{x,y}\vcenter{\hbox{\begin{tikzpicture}[anchor=base,baseline]
  \pgfmathsetmacro{\L}{4}
  \coordinate (L) at (-\L,0);
  \coordinate (R) at (\L,0);
  
  \draw[black, decorate, decoration={snake, amplitude=.4mm, segment length=2mm}]  (L) to[bend left=90,looseness=1.25] (R);
  \draw[black, decorate, decoration={snake, amplitude=.4mm, segment length=2mm}]  (L) to[bend left=75,looseness=1.20] (R);
  \draw[loosely dotted, thick] (0, {0.65*\L}) to (0, {0.55*\L});
  \node at (0.4, {(0.58*\L)}) {$q_{\mathcal{O}}$};
  \draw[black, decorate, decoration={snake, amplitude=.4mm, segment length=2mm}]  (L) to[bend left=60] (R);
  
  \draw[black, thick] (L) to[bend left=50] (R);
  \draw[black, thick] (L) to[bend left=40, looseness=1] (R);
  \draw[loosely dotted, thick] (0, {0.32*\L}) to (0, {0.22*\L});
  \node at (0.8, {(0.25*\L)}) {$q_{\Phi}$};
  \draw[black, thick] (L) to[bend left=25, looseness=0.7] (R);
  
  \node at (-\L-0.3, 0) {$x$};
  \draw[loosely dotted, thick] (0, {0.09*\L}) to (0, {-0.09*\L});
  \node at (\L+0.3, 0) {$y$};
  
  \draw[black, dashed] (L) to[bend right=40, looseness=1] (R);
  \draw[loosely dotted, thick] (0, {-0.23*\L}) to (0, {-0.33*\L});
  \node at (0.4, {-(0.29*\L)}) {$q_{\psi}$};
  \draw[black, dashed] (L) to[bend right=25, looseness=0.7] (R);
  \draw[black, dashed] (L) to[bend right=70] (R);
  \draw[black, dashed]  (L) to[bend left=90,looseness=1.25] (R);
\end{tikzpicture}}}, %
\end{equation}
where the value of $S_m$ does not matter, due to our freedom to rescale $g_m$ (but in the case of real fields is typically $S_m= \prod_\phi q^m_\phi!$). Typically, we would have had to resum the melonic insertions on each leg -- but the 2PI formalism has done this automatically.

We note that there are certain drawbacks to the 2PI reformulation of the SYK model, but these only apply at subleading orders in $N$, and so we neglect them \cite{Benedetti:2018goh}.

\subsubsection{2PI effective action for an SYK-like theory}

To illustrate this formalism, consider the 2PI effective action for an SYK-like theory: that is, for a melonic-type theory in $d=1$ with a single fermionic field $\psi_i$, we obtain %
\begin{equation}\label{eq:GammaSYK}
\frac{1}{N}\Gamma_{\text{SYK}}[G] = -\frac{1}{2}\Tr \ln G^{-1} - \frac{1}{2}\Tr C^{-1}G - \half J_m^2 \int_{t,t'} G(t,t')^q.
\end{equation}
Here:
\begin{itemize}
    \item $G(t, t') \delta_{ij}\equiv \expval{\psi_i(t) \psi_j(t')}$; so we have assumed no symmetry breaking.
    \item $N$ would be replaced by $N=M^q$ if we were studying an $\gO(M)^q$-tensor model, rather than a disorder-averaged model. All terms shown on the RHS are therefore order $N^0$.
    \item $C^{-1}(t,t') = \delta(t-t')\partial_t$ is the (matrix) inverse of the bare propagator.
    \item $J^2_m$ is the effective coupling for a complete melon\footnote{As mentioned above, we have rescaled $J_m^2$ for convenience, so it differs by: $q$ from the usual SYK normalisation, $J^2_m = J_\mathrm{SYK}^2/q$; and $q!$ from the usual Feynman-diagrammatic normalisation, $J^2_m = J_{F}^2/q!$.}.
    \item The minus signs in front of the trace terms compared to \eqref{eq:2PIeffectiveActionScalar} arise because of the fermionic character of $\psi$.
\end{itemize}
The result \eqref{eq:GammaSYK} This is well known, and directly derivable in disorder-averaged theories, via a change of variables in the replica method \cite{Benedetti:2018goh,Maldacena:2016hyu,Gross:2016kjj,Kitaev:2017awl}; it was shown in \cite{Benedetti:2018goh} that it also applies to the tensor models, though it is more complicated to derive. Adding more fermions leads to the generalized SYK model of \cite{Gross:2016kjj}; but we can now also generalize it to arbitrary dimensions, fields, and melonic interactions.

\subsubsection{2PI effective action for an arbitrary melonic theory}

Consider a melonic theory with $n_f$ fields $\{\Phi\}$, each of which has some suppressed indexed structure $\Phi_i$ associated to a melonic mechanism. Take $n_m$ melonic couplings $g_m$, so the UV perturbation is
\begin{equation}
    V=\sum_m g_m \cO_m, \quad \cO_m = \prod_\Phi \Phi^{q^m_\Phi}.
\end{equation}
By comparison with the SYK model, we can immediately write down the following $d$-dimensional 2PI effective action for $G_\Phi(x,y) \delta_{ij} \equiv \expval{\Phi_i(x)\Phi_j(y)}$.
\begin{equation}\label{eq:G2PIforArbMelonic}
\frac{1}{N}\Gamma[\{G_\Phi\}] \equiv \sum_\Phi \left(\half \Str\ln G_\Phi^{-1} +  \half \Str C^{-1}_{\Phi} G_\Phi \right) -\half\sum_m g_m^2 \int_{x,y} \prod_{\Phi} G_\Phi(x,y)^{q_{\Phi}^m}.
\end{equation}
Though its form should be clear as a generalization of \eqref{eq:GammaSYK}, we comment:
\begin{itemize}
    \item We sum over the $n_f$ dynamical fields $\Phi$, not accounting for the $N$ additional copies present due to the melonic mechanism. Thus, for the SYK or quartic tensor model, the sum only has a single contribution, $\{\Phi\} = \{\psi\}$.
    \item The supertrace $\Str X_\Phi = (-1)^{\mathrm{F}_\Phi} \Tr X_\Phi$ provides a $-1$ factor for fermionic fields ($\mathrm{F}_\psi$ is $0$ for bosons and $1$ for fermions).
    \item  All fields are assumed to be real; the generalization to complex fields requires just replacing $\half \to 1$ in the first two terms.
    \item The final term in \eqref{eq:G2PIforArbMelonic} is
    \begin{equation}\label{eq:intIs2pt}
    -\half \sum_m g_m^2\int_{x,y} \expval{\cO_m(x) \cO_m(y)},
    \end{equation}
    expanded by using the large-$N$ factorisation. This is manifestly a scalar. However, for fields in non-trivial representations of $\SO(d)\times \Gglobal$, $G_\Phi$ has indices, which must therefore be contracted; such contractions only rescale $g_m$ by constants, and so we neglect them.
    \item Since we have assumed no symmetry breaking, $R_\Phi$ then only appears in the $\Tr\mathbb{I}_R$ implicit in the first term -- so we see immediately that only the dimension of $R_\Phi$ matters.
    \item Generically, $g_m^2$ might be a homogeneous quadratic polynomial in the melonic dominant couplings, as in \cite{Prakash:2017hwq, Fraser-Taliente:2024rql}.
\end{itemize}
The UV propagator $C$ that appears in the second term plays no role in the IR, assuming $\Delta_\Phi > \Delta_\Phi^{\text{free}}$, and we drop it from $\Gamma$ hereafter: it only serves to specify the IR wedge \eqref{eq:DelGreaterThanFree}.

Naturally, the stationary point conditions $\fdv{\Gamma}{G_\Phi}=0$ are just the two-point Schwinger-Dyson equations. In fact, this is the easiest way to construct them\footnote{Of course, when performing the variation, we must remember that $G_\Phi(x,y)=(-1)^{\mathrm{F}_\Phi} G_\Phi(y,x)$ are not independent \cite{Benedetti:2018goh}.}, %
we obtain precisely the leading-$N$ two-point function SDE derived diagrammatically in \cref{app:diagrammaticProof}:
\begin{equation}\begin{aligned} \label{eq:2PIEoM} 0 = \dim R_\Phi\, G_\Phi^{-1}(x,y) + (-1)^{-\mathrm{F}_\Phi}\sum_m q^m_\Phi g_m^2 [G_\Phi(x,y)]^{q^m_\Phi -1} \prod_{\cO \neq \Phi} [G_\cO(x,y)]^{q^m_{\cO}}.
\end{aligned}\end{equation} 

\subsection{The fundamental claim proved I}

We now show how our fundamental claim arises.

In the language of \cite{Karateev:2018oml}, %
it is a standard result of CFT that a primary $\Phi$ has a unique two-point structure with an operator $\Phi^\dagger$, with scaling dimension $\Delta_\Phi$, transforming in the conjugate reflected representation $\rho^\dagger$; this representation is the complex conjugate of $\rho$ when working in Lorentzian signature. We suppress the global symmetry group and its indices. %
This two-point function is uniquely determined up to an arbitrary normalization constant $\cZ_\Phi$:
\begin{equation}\begin{aligned} \label{eq:conformalG}
G_\Phi(x,y) = \expval{\Phi_{\mu_1 \cdots \mu_s}(x) (\Phi^\dagger)^{\nu_1 \cdots \nu_s}(y)} = \cZ_\Phi [\twoPt_{\Phi}(x-y)]\indices{_{\mu_1 \cdots \mu_s}^{\mu_1 \cdots \mu_s}}\,  \mathbb{I}_{R_\Phi}.
\end{aligned}\end{equation} 
Here, $\twoPt_\Phi$ depends only on the data of the primary: the scaling dimension $\Delta_\Phi$ and the $\SO(d)$ representation $\rho$. We suppress the $G$ indices %
in the following.  

The quantity to be extremized, $\Ft$, is then defined to be the function given by the 2PI action \eqref{eq:G2PIforArbMelonic}, evaluated on the sphere with the conformal $G_\Phi$s: 
\begin{equation}\begin{aligned} \label{eq:FtEqGamma}
\Ft(\{\Delta_\Phi\}, \{\cZ_\Phi\}, \{g_m\}) \equiv -\sin(\pi d/2)\Gamma[\{G_\Phi\}]|_{S^d, \text{ conformal }G_\Phi}.
\end{aligned}\end{equation} 
It is regulated with a $-\sin \pi d/2$, just as $F$ was regulated in \eqref{eq:Ftdef}, making it demonstrably finite for all $d$. If the IR limit on the sphere is a CFT, then the constraints of conformal symmetry tell us the exact form of the $G_\Phi(x,y)$s, up to two numbers $\Delta_\Phi$ and $\cZ_\phi$. The functional extremization problem on $\Gamma$ then becomes a function extremization problem on $\Ft$. To find the sphere propagators $\twoPt_\Phi(x,y)$s, we need only conformally map the textbook two-point functions, as we did to find \cref{eq:sphereGFFpropagator}. 

Without loss of generality, we assume that all of the melons are IR relevant (if they are not, they simply disappear from the calculation entirely). Then, evaluating \eqref{eq:FtEqGamma} to leading order in $N$, we find
\newcommand{\vq}{\mathbf{q}}
\newcommand{\vG}{\mathbf{G}}
\newcommand{\mm}[1]{{\mathfrak{m}_{#1}}}
\newcommand{\Mtm}{{\tilde{\mathfrak{M}}}}
\begin{equation}\begin{aligned} \label{eq:LOfreeEnergy}
\frac{1}{N} \tilde{F}(\{\Delta_\Phi, \mathfrak{g}_m\}) =\sum_\Phi \tilde{F}_{\Phi} + \sum_m \frac{\mathfrak{g}_m}{(2R)^{2\mm{m}}} \Mtm(\mathfrak{m}_m) %
+ O\left(N^{-1}\right).
\end{aligned}\end{equation} 
We have defined the following quantities.
\begin{itemize}
    \item As in \eqref{eq:FtBosFerm}, $\Ft_\Phi$ is $\Ft$ evaluated for a generalized free field of dimension $\Delta_\Phi$ in the representation $\rhoext_\phi =\rho_\Phi \times R_\Phi$; it is independent of $\cZ_\Phi$, and linear in $\dim R_\Phi$.%
    \item The renormalized squared coupling is
    \begin{equation}\begin{aligned}
        \mathfrak{g}_m  \equiv g_m^2 \prod_\Phi \cZ_\Phi^{q^m_\Phi};
    \end{aligned}\end{equation} 
    note that in the IR the $\cZ_\Phi$s therefore appear in $\Ft$ solely through $\mathfrak{g}_m$.
    \item A convenient combination of scaling dimensions associated to each melon $m$ is
    \begin{equation}\begin{aligned}
        \mm{m} & \equiv  \sum_\Phi q^m_\Phi \Delta_\Phi-d.
    \end{aligned}\end{equation} 
    As we will see shortly, the melonic constraints are $\mm{m}=0$.
    \item The dimensionless function $\Mtm$ is proportional to the complete melon on the sphere:
    \begin{equation}\begin{aligned}
    \Mtm(\mm{m}) &\equiv [-\sin (\pi d/2)] \, \left(-\half\right) \times (2R)^{2\mm{m}} \times \int \dd^d x \dd^d y\, \sqrt{g(x)} \sqrt{g(y)} \frac{1}{s(x,y)^{2(\mm{m} +d)}}.
    \end{aligned}\end{equation}
    This complete melon integral is easily evaluated, using the homogeneity of the sphere to fix one point to zero \cite{Benedetti:2021wzt, Pufu:2016zxm, Cardy:1988cwa}: %
\begin{equation}\begin{aligned}
 \Mtm(\mm{m}) &= \frac{\pi ^d \sin \left(\frac{\pi  d}{2}\right) \Gamma \left(\frac{d}{2}\right) \Gamma \left(-\frac{d}{2}-\mm{m}\right)}{2 \Gamma (d) \Gamma (-\mm{m})}.
\end{aligned}\end{equation}  %
    This function has zeros for $\mm{m}=0, 1,2, \ldots$. 
    \item $f_\Phi^\mathrm{free}$ is the contribution from the free (UV) propagator, which is $\propto \cZ_\Phi R^{2(\Delta_\Phi^{\text{free}}- \Delta_\Phi})$. Assuming $\Delta_\Phi > \Delta_\Phi^{\text{free}}$, this will be dropped in the IR. %
\end{itemize}

The factor of $-\sin(\pi d/2)$ in the definition of $\Ft$ serves to ensure that $\Mtm'(0)$ does not have poles for even integer $d$. Since $\Ft_\Phi$ for generalized free fields is also finite, the entire $\Ft$ is indeed finite in all $d$ by construction. %

\subsection{The fundamental claim proved II}

Extremizing \eqref{eq:LOfreeEnergy} with respect to the $\Delta_\Phi$s and $\cZ_\phi$s gives
\begin{subequations}\label{eq:FtildeVariation}
\begin{align}
\forall \, \cZ_\Phi: \quad%
&\frac{1}{\cZ_\Phi}\sum_m q^m_\Phi \mathfrak{g}_m\Mtm\left(\mm{m}\right)(2R)^{-2\mm{m}} = 0,\label{eq:ZphiVariation}\\
\forall \, \Delta_\Phi: %
\quad&  \odv{\tilde{F}_\Phi}{\Delta_\Phi} + \sum_m q^m_\Phi \mathfrak{g}_m \Mtm'\left(\mm{m}\right) (2R)^{-2\mm{m}} =0, \label{eq:deltaPhiVariation}%
\end{align}\end{subequations}
where we have immediately used \cref{eq:ZphiVariation} to cancel the term $\propto \log R$ that otherwise would appear in \cref{eq:deltaPhiVariation}, and dropped the $N$-subleading terms. %

The extrema of $\Ft$ satisfying $\mm{m}=0$ for all $m$ correspond to the extrema of the functional $\Gamma$ that are independent of $R$, and therefore can be consistently mapped to flat space. There exist other solutions to \eqref{eq:ZphiVariation}, but they lead to terms of different order in $R$ in \cref{eq:deltaPhiVariation}; the dependence of the solutions on $R$ implies that we do not have a consistent IR solution (which should be $R$-independent for large $R$). Contributions from any $g_m$s with $\mm{m} >0$ will not survive the IR limit, and therefore the associated melonic constraint will not be applied. As mentioned above, we will typically neglect this possibility.

To make contact with the usual analysis of the full Schwinger-Dyson equations, it is clear that any such solutions with $R$-dependence do not satisfy \eqref{eq:2PIEoM}; that is, they are not extrema of the full $\Gamma[G]$; they are only extrema of the conformal slice of $\Gamma$, which we have defined to be $\Ft(\Delta_\phi, Z_\Phi)$. It is only the $R$-independent extrema of $\Ft$ that give $G$s that extremize $\Gamma[G]|_{S^d}$. 
We can therefore expand \eqref{eq:LOfreeEnergy} around $\mm{m}=0$, using
\begin{equation}\begin{aligned}\label{eq:MtmExpanded}
\Mtm(\mm{m}) (2R)^{-2 \mm{m}} = \frac{\pi^{d+1}}{\Gamma(d+1)} \mm{m} + O(\mm{m}^2).
\end{aligned}\end{equation} 

Substituting \eqref{eq:MtmExpanded} into \eqref{eq:deltaPhiVariation}, we see that the functional extremization problem is, in the IR, equivalent to extremizing the function
\begin{equation}\begin{aligned}
\frac{1}{N}\Ft(\{\Delta_\Phi, \mathfrak{g}_m'\})  = \sum_\Phi \Ft_\Phi + \sum_m \mathfrak{g}'_m \left(\sum_\Phi q^m_\Phi \Delta_\Phi-d\right) \
\end{aligned}\end{equation}  
with respect to the Lagrange multipliers
\begin{equation}\begin{aligned}
\mathfrak{g}_m' = \frac{\pi^{d+1}}{\Gamma(d+1)} g_m^2 \prod_\Phi \cZ_\Phi^{q^m_\Phi}.
\end{aligned}\end{equation} 
Hence, the extremum of $\Ft(\{\Delta\})$ corresponds precisely to the actual value of $\Ft_\mathrm{CFT}$ for this CFT, and is just
\begin{equation}\begin{aligned}
\Ft_\mathrm{CFT} = \sum_\Phi \Ft_\Phi|_{\Delta_\Phi}.
\end{aligned}\end{equation} 
The complete melon has been regulated to zero; this can be understood as the fact that the trace of the identity is zero in continuous dimension, $\Tr \id =0$ \cite{Benedetti:2021wzt}.
Naturally, all of this agrees with the equations obtained by a diagrammatic analysis in the appendix, \eqref{eq:melonicsummary}. 

This is the fundamental claim \eqref{eq:FmaxSummary}: the melonic interaction precisely implements the (linear) melonic constraint in the $\tilde{F}$-extremization procedure. Large-$N$ factorisation immediately gives (see \eqref{eq:intIs2pt}) that the IR scaling dimension of each of the monomials $\cO_m$ in $V$ is $\mm{m}+d$, so we see that the melonic constraints ensure that the interaction term is marginal in the IR. This is consistent, as the large-$N$ limit protects the form of the $V$, so no additional terms are generated. %
This is precisely the same as the supersymmetric case. 

Thus, to conclude in words: the conformal IR solution of a melonic QFT can be found by extremizing the sum of the free energies of a collection of generalized free fields, subject to a marginality constraint $\mm{m}=0$ (or $\mathfrak{g}_m=0$) for each melon. Since any non-Gaussian behaviour in the correlators is $N$-subleading, the melonic CFTs are therefore precisely the mean field theories with constrained extremal $\Ft$.

\section{The pattern: large-\texorpdfstring{$n$}{n} vector models} \label{sec:vectorModels}

The leading order of the large-$n$ vector models is melonic; we can therefore use \eqref{eq:FmaxSummary} to solve for the critical CFT$_d$. Certain features of the melonic models are already visible here in a more familiar context: namely, the multiple vacua solutions; the collision and hence disappearance (complexification) of these vacua as $d$ is varied; and the missing solutions at certain integer values of $d$. We will however promote the vector model to a model that is melonic for all values of $n$; thus by varying $n$, we interpolate from a standard large-$n$ vector theory to a more typical melonic theory.%

\subsection{The vector model and the disordered vector model}

We study the $\gO(n)$ $(\phi_I \phi_I)^2$ vector model \cite{Vasiliev:1981dg, Goykhman:2019kcj}. %
Introducing an auxiliary field $\sigma$, the Lagrangian is given by 
\begin{equation}\begin{aligned} \label{eq:vectorOn}
\cL_{O(n)} = \sum_{I=1}^n \half \phi\indices{_I} (-\partial^2) \phi_{I}  - \frac{1}{4} (\sigma \sigma) + \sum_{I=1}^n g \sigma\phi\indices{_I} \phi_{I}.
\end{aligned}\end{equation} 
The leading propagator corrections are (with explicit symmetry factors)
\begin{equation}\begin{aligned}
\Pi_\phi = \verticalcenter{\begin{tikzpicture}\begin{feynman}
  \vertex (a)  at (0,0);
  \vertex (b) at (1,0);
  \vertex (c) at (2,0);
  \vertex (d) at (3,0);
  \diagram* {
    (a) -- [scalar] (b),
    (b) -- [boson, half left, looseness=1.5] (c),
    (c) -- [scalar, half left, looseness=1.5] (b),
    (c) -- [scalar] (d)
  };\end{feynman}\end{tikzpicture}} + O(n^{-1}), \quad 
\Pi_\sigma= \frac{n}{2} \, \verticalcenter{\begin{tikzpicture}\begin{feynman}
  \vertex (a) at (0,0);
  \vertex (b) at (1,0);
  \vertex (c) at (2,0);
  \vertex (d) at (3,0);
  \diagram* {
    (a) -- [boson] (b),
    (b) -- [scalar, half left, looseness=1.5] (c),
    (c) -- [scalar, half left, looseness=1.5] (b),
    (c) -- [boson] (d)
  };\end{feynman}\end{tikzpicture}} + O(n^0) \label{eq:vectorModelLO}
\end{aligned}\end{equation} 
which are indeed melons, and hence the theory to leading order in $n$ is melonic, with melonic data
\begin{equation}\begin{aligned} \label{eq:OnMelonicData}
q_\phi =2, \quad q_\sigma = 1, \quad \frac{\dim \rhoext_\phi}{\dim \rhoext_\sigma} = n.
\end{aligned}\end{equation}
To aid our later comparison with the melonic models, we want to make this theory exactly melonic for all values of $n$. We can achieve this by adding a melonic mechanism such as a disordered coupling constant\footnote{As usual, a suitable generalization of the Amit-Rogansky model \cite{Benedetti:2020iku}, or a tensor model, would also give the same results without the disorder average.} $g_{aij}$, just as was done in \cite{Chang:2021wbx,Shen:2023srk}. Promoting both $\phi_I$ and $\sigma$ to $N$-component fields, \eqref{eq:vectorOn} becomes 
\begin{equation}\begin{aligned}
\cL_{\text{disordered }O(n)} = \sum_{i=1}^N \sum_{I=1}^n \half \phi\indices{^i_I} (-\partial^2) \phi\indices{^i_I}  - \sum_{a=1}^N\frac{1}{4} (\sigma^a \sigma^a) + \sum_{a,i,j=1}^N  \sum_{I=1}^n g_{aij} \sigma^a \phi\indices{^{i}_I} \phi\indices{^j_I},
\end{aligned} \end{equation}
where we disorder average the coupling $g_{aij}$. We then take $N\to \infty$, while keeping $n$ finite. The melonic data \eqref{eq:OnMelonicData} is identical, and so $2\Delta_\phi + \Delta_\sigma = d$ holds as an exact statement for finite $n$ as well, up to leading order in $N$. In this sense, this disordered vector theory is a consistent extension for finite $n$ of the leading large-$n$ physics of the vector model. 

Assuming as usual unbroken symmetry, and using \eqref{eq:FmaxSummary}, we find that in the IR of the disordered vector model, to leading order in the melonic mechanism's $N$, but exactly in $n$,%
\begin{equation}\begin{aligned} \label{eq:polesForVector}
\frac{\Gamma (2 \Delta_\phi ) \Gamma (d-2 \Delta_\phi ) \Gamma \left(\frac{d}{2}-\Delta_\phi \right) \Gamma \left(\Delta_\phi -\frac{d}{2}\right)}{\Gamma (\Delta_\phi ) \Gamma \left(\frac{d}{2}-2 \Delta_\phi \right) \Gamma (d-\Delta_\phi ) \Gamma \left(2 \Delta_\phi -\frac{d}{2}\right)} = \frac{n}{2} \text{, where } 2\Delta_\phi + \Delta_\sigma = d, 
\end{aligned}\end{equation}  
with the IR wedge defined by $\Delta_\phi > \frac{d-2}{2}, \, \Delta_\sigma > \frac{d}{2}$.

We now make the connection to the standard large-$n$ vector model by \textit{also} taking the large-$n$ limit. This has a solution for $\Delta_\phi = \frac{d-2}{2} + \gamma_\phi$, $\Delta_\sigma =2 + O(1/n)$, with
\begin{equation}\begin{aligned}
\gamma_\phi = \frac{1}{n} \frac{2 (4-d) \Gamma (d-2)}{d\, \Gamma(2-\frac{d}{2}) \Gamma (\frac{d}{2}-1)^2 \Gamma (\frac{d}{2})} + O\left(\frac{1}{n^2}\right), %
\end{aligned}\end{equation} 

which is exactly the standard result for the anomalous dimension of the field $\phi_I$ \cite{zinn-justin_quantum_2002}. For $2<d<4$ it is positive, and so lies in the IR wedge.

There is one slight complication: at this stage, the correction to $\Delta_\sigma$ in the large-$n$ limit for the original vector model is undetermined, as there $\Delta_\sigma = d-2\Delta_\phi$ only up to $O(1/n)$ corrections. However, the actual value of $\Delta_\sigma$ would be calculable by studying the spectrum of bilinears appearing in the OPE $\phi \times \phi$. Nonetheless, we have the result described in \cite{Chang:2021wbx}, of a one-parameter family that interpolates from the critical $\gO(n)$ model (dual to higher spin AdS gravity) at $n\to\infty$ to the theory with a classical string dual. %

\subsection{Further solutions}

\begin{figure}[ht]
\centering
    \includegraphics[width=0.8\textwidth]{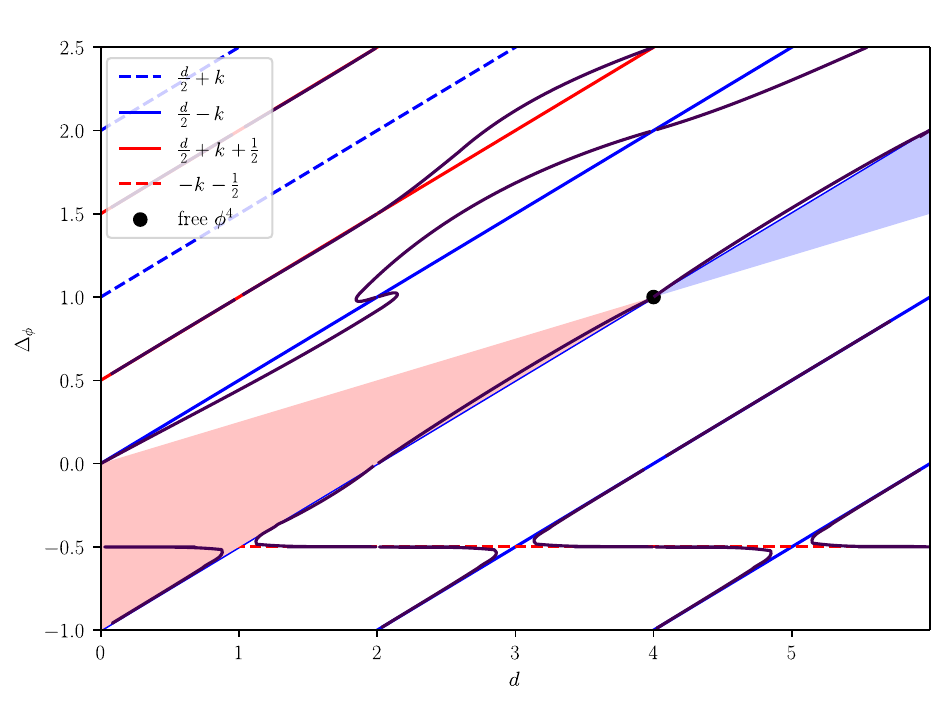}
    \caption{Large (but finite) $n$ contours superimposed in black onto the poles of \eqref{eq:polesForVector}. Note the absence of the $\Delta_\phi \sim \frac{d}{2} + k, k>0$ contours; this is due to the complexity of the anomalous dimensions around those poles. The curious twisted behaviour of the $\Delta_\phi \sim \frac{d}{2}$ line is due to the triple pole. The line descending from the free theory should be in the IR consistent (red) region only for $2<d<4$. We have taken $n=200$, but note that the size of the perturbation around $\Delta_\phi = \frac{d-2}{2}$ has been manually enhanced for clarity, to indicate in which dimensional range the scaling dimensions lie in the regions of IR and UV validity.}
    \label{fig:otherBranchesVectorModelFinite}
\end{figure}

The story above is the usual solution of the $\gO(n)$ vector model, but in fact \cref{eq:polesForVector} has many other solutions for $\Delta_\phi$ in generic $d$. These are shown in \cref{fig:otherBranchesVectorModelFinite} by the black contours. Note that for $d>1$, only one of the contours lies within the IR wedge \eqref{eq:DelGreaterThanFree}, shown in red (for completeness we also show the UV wedge in blue). However, we are always allowed to modify the free propagator to that of a GFF -- that is, some formal expression like $\phi (-\partial^2)^\zeta \phi$ for any $\zeta$ -- usually at the cost of unitarity and locality \cite{Benedetti:2021wzt}. 
Essentially, this modifies the UV scaling dimension $\Delta^{\mathrm{free}}_{\phi,\sigma}$, which changes the IR wedge. Then, the modified Lagrangian can access these extra solutions in its conformal limit.

Let us understand where these additional solutions come from in the large-$n$ limit; these lessons will transfer to the generic melonic case. The gamma function has no zeros. Therefore, in the large-$n$ limit, all solutions to \eqref{eq:polesForVector} must come from near the poles $\Delta^0_\phi$ of the gamma functions in the numerator. Of course, these poles must not coincide with poles in the denominator; expanding about them with $\Delta_\phi \equiv \Delta_\phi^0 + \gamma_\phi$, we can determine the leading-order anomalous dimensions for the fixed points. The order in $n$ of the correction then depends on the multiplicity of the pole. A pole of multiplicity $m$ at some $\Delta_\phi^0(d)$ leads to a leading-order equation
\begin{equation}\begin{aligned}
\frac{f(d)}{\gamma_\phi^m} +O(n^0) = \frac{n}{2} \quad \implies \quad \gamma_\phi \propto \frac{1}{n^{\frac{1}{m}}};
\end{aligned}\end{equation} 
this may be complex, depending on $m>1$ and the sign of $f(d)$ (which is always real, as $\Gamma(x)$ is real for real $x$). Thus, in practice, some of the perturbative poles will not give physical IR CFTs. 
The locations and multiplicities of such poles for integer $k$ are
\begin{equation*}\begin{aligned}
\begin{matrix}
     & \textbf{pole multiplicity} & \Delta_\phi^0 & \text{ values } & \gamma_\phi \textbf{ values}\\ \hline
    \text{(i)} &  \textbf{triple pole }&\frac{d}{2}& \frac{d}{2} & \text{ one real, two imaginary } \propto \frac{1}{n^{1/3}} \\
     \text{(ii)} & \text{double pole }& \frac{d}{2} + k, k>0 &\quad \frac{d}{2} + 1,\frac{d}{2} + 2,\ldots & \text{ for d<4, two imaginary } \propto \frac{i}{n^{1/2}} \\
     \text{(iii)} & \text{single pole }&\frac{d}{2} - k, k>0&\quad \frac{d}{2}, \frac{d}{2}- 1,\ldots & \text{ real } \propto \frac{1}{n}\\
     \text{(iv)} & \text{single pole }&\frac{d}{2} + k + \half, k \ge 0&\quad \frac{d}{2} + \half \frac{d}{2} + \frac{3}{2},\ldots & \text{ real } \propto \frac{1}{n}\\
     \text{(v)} & \text{single pole }&-k -\half, k \ge 0 &\quad -\half, \, -\frac{3}{2}, \ldots & \text{ real } \propto \frac{1}{n}
\end{matrix}
\end{aligned}\end{equation*} 
These five cases can be written compactly as 
\begin{equation}
    \text{(i-iii) } \Delta^0_\phi = \frac{d}{2} \pm k \text{ and (iv-v) } \Delta^0_\sigma = \frac{d}{2} \pm \left(\frac{d}{2} +2 k +1 \right), \text{ for } k \ge 0.
\end{equation}%
To make it clear that the solutions are perturbations around the pole lines, they are shown in colour in \cref{fig:otherBranchesVectorModelFinite}. %
We make the following comments:
\begin{itemize}
    \item To ensure similarity to the actual large-$n$ vector model, we have chosen $n=200$ for the ratio of degrees of freedom of $\phi$ and $\sigma$. This ratio is tunable, and, as shown in the tables, all anomalous dimensions are order $1/n^k$. For small $n$, the solutions depend strongly on $n$.
    \item For example, we have one real and two complex solutions around $\Delta_\phi =\frac{d}{2}$:
\begin{equation}\begin{aligned}
\Delta_\phi = \frac{d}{2} + \frac{\alpha}{n^{1/3}} + O(1/n^{2/3}), \quad \alpha^3 = \frac{\Gamma (d)}{\Gamma \left(-\frac{d}{2}\right) \Gamma \left(\frac{d}{2}\right)^3},
\end{aligned}\end{equation} 
which gives the notable twisted structure.
\item There are no visible solutions for $\Delta^0_\phi = \frac{d}{2} + k$, $k=1,2$, since they are complex.
\item We note that for $d=1$ exactly, for $\Delta_\phi>0$, there is only one real solution, with $\Delta_\phi = \frac{d}{2} + O(1/n^{1/3})|_{d=1}$. %
However, there are an infinite number of real solutions perturbatively in $d=1+\epsilon$, which is why the contours appear unbroken. %
\item The fact that $\gamma_\phi$ is positive for $2<d<4$ means that the $\Delta_\phi = \frac{d-2}{2} + \gamma_\phi$ solution is consistent in this region. We observe the presence of a second consistent IR solution (that is, one lying within the IR wedge) for $\Delta_\phi = -\half + O(1/n)$ for $d<1$, and hence $\Delta_\sigma = d+1 + O(1/n)$, that seemingly descends from the free $d=1$ boson. %
\item We can immediately identify the poles at $\Delta^0_\phi = \frac{d}{2} - k$ in the large-$n$ limit as being the real $\Box^{k}$ CFT, which is dual to the minimal type-A$_k$ higher spin gravity \cite{Sun:2020ame,Bekaert:2013zya, Brust:2016zns}.%
\end{itemize}

This analysis extends immediately to other large-$n$ vector models, such as the Gross-Neveu model \cite{Zinn-Justin:1991ksq}, where we would take
\begin{equation}\begin{aligned}
\frac{\dim \rhoext_\psi}{\dim \rhoext_\sigma} =2 \Tr[\mathbb{I}_s]\times n,
\end{aligned}\end{equation} 
and again solve to leading order in $n$. The result matches the computations of \cite{Zinn-Justin:1991ksq,Goykhman:2020tsk}, and %
changing the $\dim \rhoext$s, also solves the chiral and non-abelian \cite{Gracey:2018qba,Gracey:2021yyl} extensions. Hence, we have a simple example of how, as promised, only the dimension of complicated finite symmetry representations matters. %

We have shown that the disorder-averaged vector model indeed gives a consistent extension of the leading-$n$ physics of the critical vector model ($n\to \infty$) to finite values of $n$. Of course, the melonic structure of the standard large-$n$ vector model does not persist to the next order in $n$: if we attempt to compute the subleading terms in $\Ft$, along the lines of \cite{Tarnopolsky:2016vvd}, we find non-melonic diagrams. This means that when solving for the IR, we no longer have the neat interpretation of the constrained extremization of $\Ft(\Delta_\phi,\Delta_\sigma)$ for two GFFs, $=N\Ft_b(\Delta_\phi) + \Ft_b(\Delta_\sigma)$. 

The large-$n$ vector models provide simple examples in which we can observe these characteristic features. We turn next to the melonic models, in which case the equivalent of the parameter $n$ is decreased to order one, and so the anomalous dimensions are also order one. This makes it harder to see directly the reason for, for example, the disappearing contour lines (the generation of complex anomalous dimensions), but they occur for reasons identical to those shown here.%

\section{Melonic models: some examples} \label{sec:melonicModels}

For a generic melonic model, the anomalous dimensions are order one, and so it can be harder to interpret what happens as we change $d$ -- this is why we began with the large-$n$ vector models in \cref{sec:vectorModels}. In this section, we demonstrate how the characteristic features identified in the vector models also hold here:
\begin{enumerate}
    \item There are multiple conformal vacua, only some of which lie inside the IR wedge, for theories with $n_f>n_m$. As in the vector case, each line arises from a perturbation around the pole of a gamma function.
    \item At certain values of $d$, pairs of real solutions collide and become a complex conjugate imaginary pair: thus we have disappearing lines of solutions.
    \item There are gaps in the solution contours at certain integer values of dimension $d$. 
    \item The existence and location of real solutions depends strongly on the solutions on the ratios of $\dim\rhoext_\phi$s.
\end{enumerate}
We use three multi-field models as our example: first, two single-interaction models, and then a multi-interaction model, which also possesses supersymmetric vacua. We do not discuss the single-field melonic models (or more generally, the theories with $n_f=n_m$), as they have only a single solution, $\Delta = d/q$, directly from the constraint(s).

\subsection{Two fields, one interaction: the quartic Yukawa model} \label{sec:QuarticYukawaFull}
\begin{figure}
    \centering
    \includegraphics[width=0.6\linewidth]{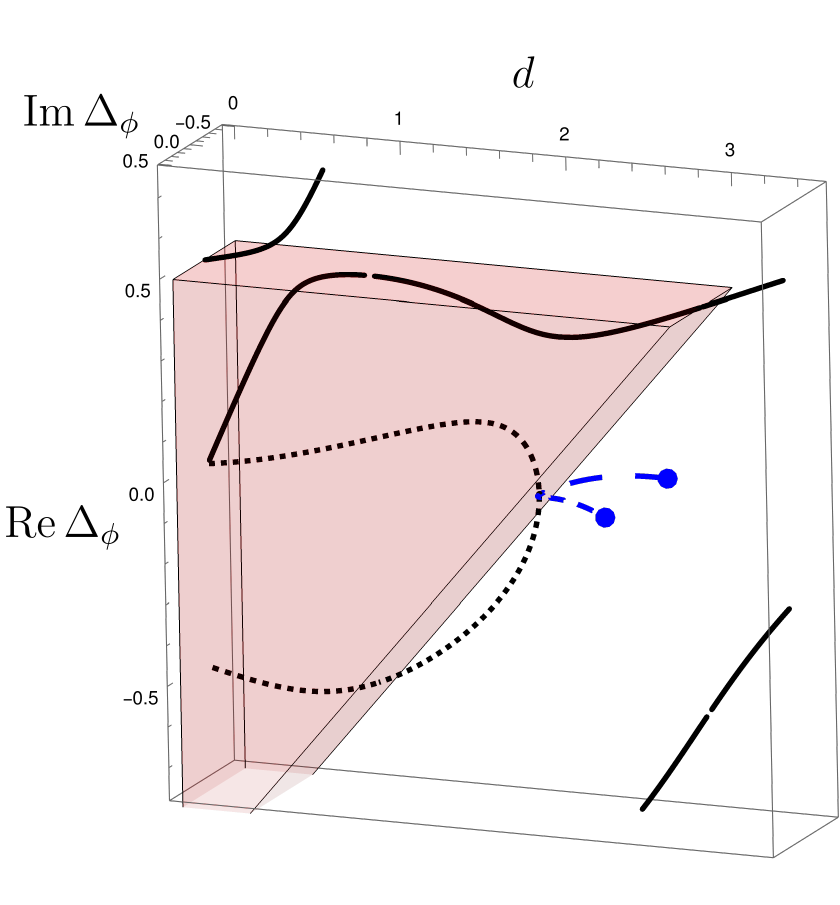}
    \caption{Plot of solutions for the scaling dimension $\Delta_\phi$ in the $\phi^2 \bar\psi \psi$ melonic model. $\Delta_\psi$ is obtained from the melonic constraint $2\Delta_\phi + 2\Delta_\psi=d$. Complex solutions are blue and dashed, and leave the volume of the plot at the blue dots. The real slice of this figure corresponds to \cref{fig:phi2psibarpsiIRwedge}. On the real axis, solid lines represent maxima of $\Ft$; dashed lines represent minima. The gaps in the contours are real, and indicate missing solutions in $d=1$ and $d=3$ here.} %
    \label{fig:QuarticYukawaSolutionPlot3D}
\end{figure}

In \cref{fig:QuarticYukawaSolutionPlot3D}, we give the contour plot of the solution space for the $\phi^2 \bar\psi \psi$ melonic model, tuned so that
\begin{equation}\begin{aligned} \label{eq:QYdimrat}
\frac{\dim \rhoext_\psi}{\dim \rhoext_\phi} = 2 \Tr \spinid = 4.
\end{aligned}\end{equation} 
Unlike \cref{fig:phi2psibarpsiIRwedge}, we also include here the (blue, dashed) complex solutions; these collide at $d=2$ on the free scalar line, giving the two real solutions that exist for $d \le 2$. The details of the contours and the occurrence of complexification occurs depend strongly on the ratio \eqref{eq:QYdimrat}, as was demonstrated in \cite{Fraser-Taliente:2024rql}.

\subsection{Three fields, one interaction: the Popovi\'c model}

\begin{figure}
    \centering
    \begin{subfigure}{0.65\textwidth}
    \includegraphics[width=\textwidth]{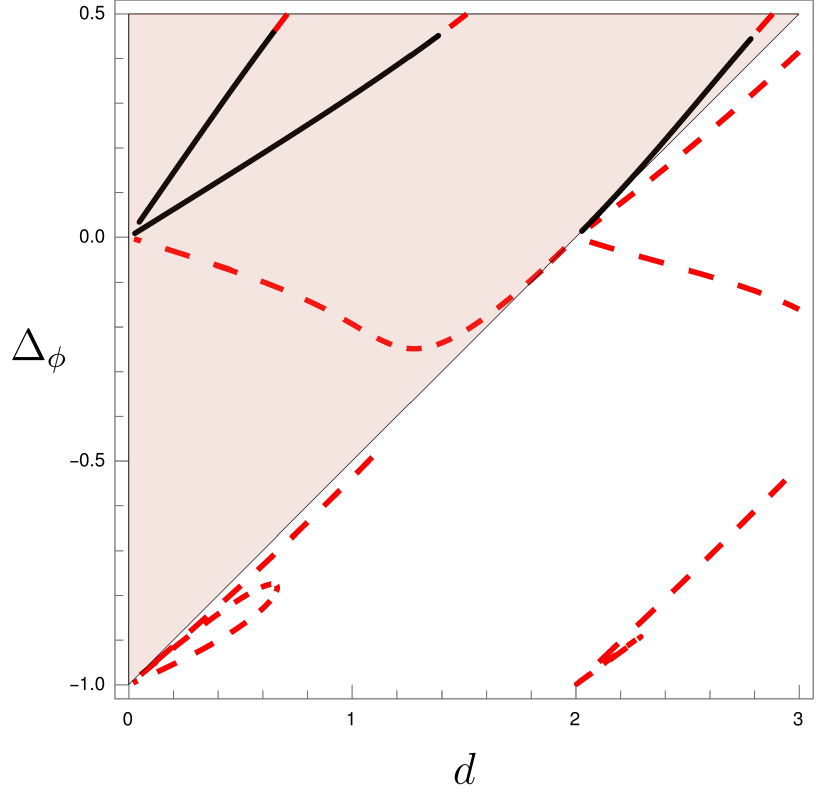}
    \caption{Contour plot of $\Delta_\phi$ against $d$.} %
    \label{fig:PopN5r2}
    \end{subfigure}%
    \vspace*{0.3cm}
    \begin{subfigure}{0.65\textwidth}
    \vspace*{\fill}
    \includegraphics[width=\textwidth]{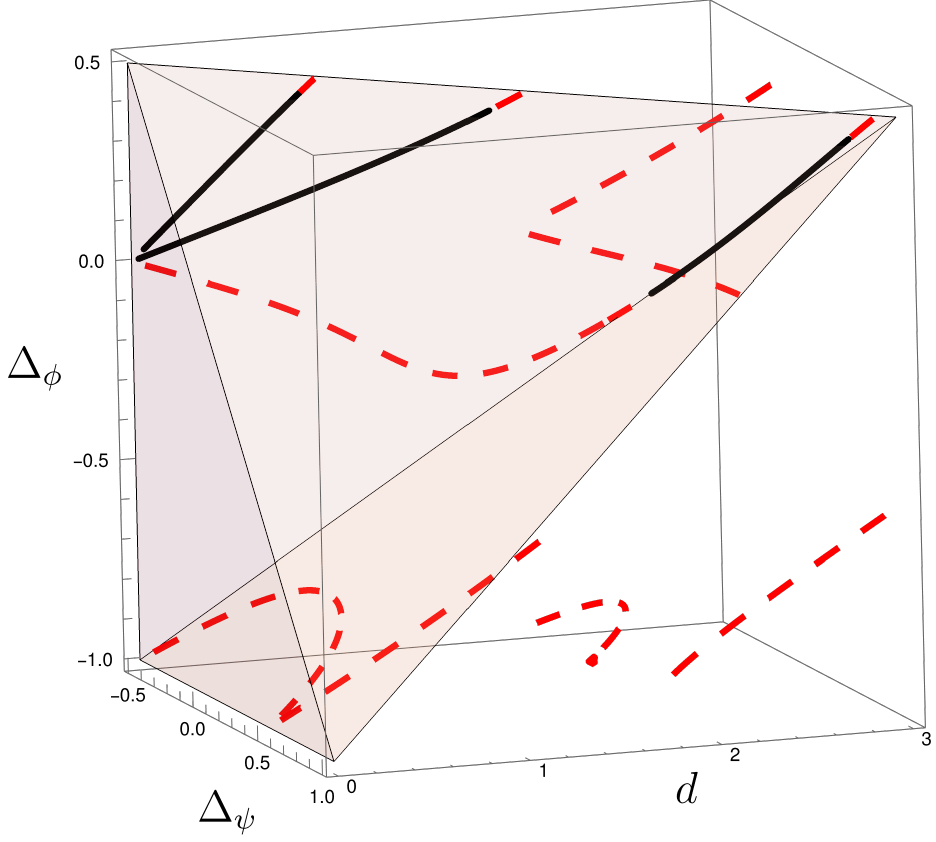}
    \vspace*{\fill}
    \caption{Angle view, showing 3d structure.}
    \end{subfigure}
    \caption{Scaling dimension solutions of the melonic Popovi\'c model \cref{eq:melPopLag} for various $d$, illustrated for $n=5$, $\Tr[\spinid]=2$; it demonstrates a complex network of conformal vacua. The IR wedge is now an IR tetrahedron, which is shaded: black lines lie inside it; the dashed red lines lie outside. Unlike before, we do not indicate the maxima and minima. The rightmost line in \cref{fig:PopN5r2} corresponds to the solution $\Delta_\phi = \frac{d-2}{2} + O(1/n), \Delta_\psi = \frac{d-1}{2} + O(1/n)$. The left-hand lines are the $\Delta_\phi = \frac{d}{2} \pm O(1/n^\half), \Delta_\psi = \frac{d-1}{2} + O(1/n)$ solutions.} 
    \label{fig:Popoviccontours}
\end{figure}

In the case of more fields, we find a higher-dimensional generalization of the above behaviour. In \cref{fig:Popoviccontours}, we show the IR solutions for a melonic version of the Popovi\'c model \cite{Popovic:1977cq}. With melonic mechanism suppressed, the Lagrangian for this is
\begin{subequations}\label{eq:melPopLag}
\begin{equation}\begin{aligned}
\cL_\text{Popovi\'c} \sim \bar\phi_I (-\partial^2) \phi_I  + \bar{\psi}_I(-\slashed{\partial}) \psi_I - \bar\chi \chi + g_0(\bar\chi \phi_I \psi_I  + \bar\psi_I \phi^\dagger_I \chi),
\end{aligned}\end{equation} 
where $I$ is summed from $1$ to $n$, and $\chi$ is a fermionic auxiliary field\footnote{Note that the Popovi\'c model is essentially an $\gO(n)$ vector model, where the singlet field is $\chi \propto \phi_I \psi_I$ instead of the usual $\phi_I \phi_I$.}, which becomes dynamical. We choose arbitrarily
\begin{equation}\begin{aligned}
\frac{\dim \rhoext_\psi}{\dim \rhoext_\phi} = \Tr[\spinid]= 2,\quad \frac{\dim \rhoext_\phi}{\dim \rhoext_\chi} = n = 5.
\end{aligned}\end{equation} 
\end{subequations}
The melonic mechanism constrains $\Delta_\phi + \Delta_\psi + \Delta_\chi =d$, and so eliminating $\Delta_\chi$, the extrema of $\Ft$ lie in the three-dimensional space $(d,\Delta_\phi,\Delta_\psi)$ shown in the figure.
The IR wedge becomes an IR tetrahedron
\begin{equation}\begin{aligned}
\Delta_\phi > \frac{d-2}{2},\quad \Delta_\psi > \frac{d-1}{2}, \quad \Delta_\chi > \frac{d}{2}.
\end{aligned}\end{equation} 
We still find a discrete set of vacua in each $d$, only some of which, drawn in black, lie within the IR tetrahedron.

\subsection{Multi-interaction melonics; a supersymmetric model} \label{sec:SUSYmelons}

As described above, we can consider multiple melonic interactions, i.e. $n_m>1$.
One natural case where these arise is in the supersymmetric melonic theory of a single chiral field. 
Recall that the standard supersymmetric $\Ft$-extremization trivially collides with the melonic $\Ft$-extremization.
The advantage of the melonic theory is that we are also permitted to consider the vacua with broken SUSY.
We now provide a somewhat shortened version of the discussion of section 4.5.1 of \cite{Fraser-Taliente:2024rql}.
We take the melonic-type theory of a single complex scalar superfield with four supercharges, as given in \cite{Lettera:2020uay}.
As usual, we suppress the melonic mechanisms, which can be found in \cite{Murugan:2017eto,Chang:2018sve,Popov:2019nja,Lettera:2020uay}.
The form of the superpotential makes the $\Ft$-extremization trivial,
\begin{equation}\begin{aligned} \label{eq:QYauxSUSYSol}
W[\Phi] \sim g \Phi^4, \quad  \Delta_\Phi=\frac{d-1}{4}
\end{aligned}\end{equation} 
However, we can also consider breaking up the superfield into its components in the usual way, 
\begin{equation}\begin{aligned}
\Phi= \phi + \theta \psi - \theta^2 X. %
\end{aligned}\end{equation} 
In this case, %
we obtain a potential of schematic form
\begin{equation}\label{eq:susyBrokenV}
V(\phi, \psi, X) = \rho (X \phi^3 + X^\dagger (\phi^\dagger)^3) + \lambda \phi^\dagger \phi \bar\psi \psi,
\end{equation}
where supersymmetry gives the relation $\rho \sim \lambda$. This is the full quartic Yukawa model, with auxiliary field, of \cite{Fraser-Taliente:2024rql}, which was defined as $h\lambda_\text{prismatic}$ there. 

\begin{figure}
    \centering
    \includegraphics[width=0.8\linewidth]{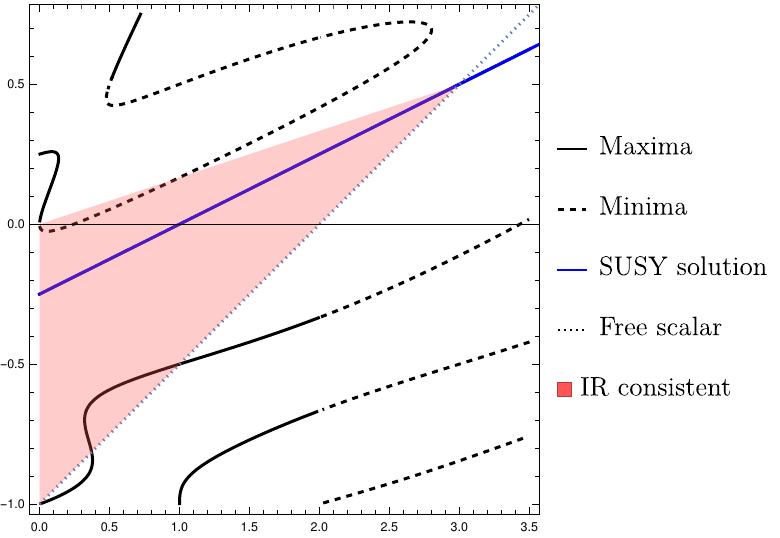}
    \caption{The supersymmetry-breaking vacua of the theory of a single complex scalar superfield with four supercharges; i.e. with potential \cref{eq:susyBrokenV}. The SUSY-preserving solution \eqref{eq:QYauxSUSYSol} is the line $\Delta = \frac{d-1}{4}$.} %
    \label{fig:hlamprismatic-r2}
\end{figure}

Let us now consider SUSY-breaking vacua, for which we allow $\rho$ and $\lambda$ to vary independently. We take $\dim \rhoext_X= \dim \rhoext_\phi$ and $\dim \rhoext_\psi = 2 \dim \rhoext_\phi$. 
Since we have two interactions, there are now two melonic constraints,
\begin{equation}\begin{aligned}
\Delta_X + 3 \Delta_\phi = d \, \text{ and } \, 2\Delta_\phi + 2\Delta_\psi = d,
\end{aligned}\end{equation} 
and so the extrema are only in the space $(d,\Delta_\phi)$, which we plot in \cref{fig:hlamprismatic-r2}. The supersymmetric solution \eqref{eq:QYauxSUSYSol}, where $\Delta_\phi = \Delta_\psi - \thalf= \Delta_X-1$, is the straight line in the figure. In addition, the component melonic approach has given non-perturbative access to all the SUSY-breaking vacua; however, in this case the additional vacua are only IR consistent for $d<1$.

\section{Discussion} \label{sec:Discussion}

In this paper we have established our fundamental claim \eqref{eq:FmaxSummary}; that is, the melonic CFTs in the strict large-$N$ limit are precisely a mean field theory, or collection of generalized free fields, with extremal sphere free energy -- subject to constraints that correspond to IR marginality of the potential. This holds regardless of any finite symmetry groups and the interaction structure: hence we have a complete solution and classification of melonic CFTs. Now, we expected mean field behaviour in the large-$N$ limit, due to factorization. However, since $\Ft$ is also thought to count the number of degrees of freedom in a CFT, this has the pleasing interpretation that we extremize the number of IR degrees of freedom.

To establish this claim, we used the fact that the quantum solution of a field theory lies at the extremum of the two-particle effective action $\Gamma_\mathrm{2PI}$. This holds even when $\Gamma_\mathrm{2PI}$ is evaluated on the IR conformal slice of solutions; thus, the quantum solution lies at an extremum with respect to the trial scaling dimensions of the fundamental fields, which leads directly to a non-perturbative definition of $\Ft(\{\Delta_\phi\})$. We then illustrated this procedure and its results using various example melonic CFTs, and saw that much of the IR structure can be understood by generalizing that of the large-$N$ vector models (including the $\Box^k$ CFTs).

Future questions fall into two categories. The first concerns the nature of the IR structures that arise here: 
\begin{enumerate}
\item We often find a discrete choice of vacua in the IR, some of which are maxima, some of which are minima, and some of which are saddle points. %
Can we identify a mechanism to select which one is physically realised? Equivalently, in the SUSY mechanism, unitarity requires that $F$ and $a$ must be maximized; is the same true here?
Of course, these theories can only be unitary in integer dimensions, due to the presence of evanescent operators \cite{Hogervorst:2015akt,Ji:2018yaf}; nonetheless, is there a sense in which $\Ft$-maximization holds for QFTs that are the analytic continuations in $d$ of unitary theories?
\item Along similar lines: is $\Ft_\mathrm{IR} \le \Ft_{UV}$ always true, at least within the IR wedge? This would give further hints towards a generalized $\Ft$-theorem, valid in continuous dimension \cite{Giombi:2014xxa,Fei:2015oha}. The fact that the usual free field values are local maxima of $\Ft_\phi$ almost -- but not quite -- shows this, as an unconstrained maximum always gives an upper bound for a constrained maximum. %
\item The additional vacua lying outside the unitarity wedge can be understood in the case of the vector model as corresponding to the $\Box^k$ CFTs -- these also have known AdS duals \cite{Sun:2020ame,Bekaert:2013zya, Brust:2016zns}. It would be nice to understand the additional vacua in other melonic models to the same degree.
\item In certain integer dimensions $d$, we find no solutions for the field scaling dimensions, despite the existence of perturbative solutions around that $d$ (this is just as in \cite{Biggs:2023mfn}). What, therefore, happens in the IR of these theories? \cite{Schaub:2024rnl}
\item We can also compute the scaling dimensions of operators in the OPE of the fundamental fields, as well as their OPE coefficients, at least for the tensorial realisation of the SYK model \cite{Benedetti:2021wzt}; what can we discover by analysing these? Various divergences are evident in the scaling dimensions \cite{Giombi:2018qgp,Fraser-Taliente:2024rql} in the case when the scaling dimension of one of the fields hits zero. Does the $\Ft$-extremization perspective aid in understanding the true fate of these theories in the IR, perhaps as logarithmic CFTs?
\item The melonic and supersymmetric CFT mechanisms are identical in practice. The only difference is that for the latter, at finite $N$ the Lagrange multiplier is a more complicated function of the coupling constant, i.e. $\lambda \sim g^2 \prod_\phi \cZ_\phi^{q_\phi} + O(g^3)$. This coincidence arises  because the form of the potential is protected in both, by supersymmetry and large-$N$ respectively; hence only field renormalization occurs. However, is there more to this than coincidence? %
In any case, we can directly import the results from $F$ and $a$-maximisation to the melonic SCFTs. This may lead to a neat explanation of their various spectral divergences: for example, in \cite{Popov:2019nja}, we obtain $\Delta =0$ in  $d=1$, and therefore a missing tower of operators in (at least) the $BB$-type bilinears; likewise in \cite{Lettera:2020uay}. 
\end{enumerate}
The second category involves extending this procedure in various directions:
\begin{enumerate}
\item It is clear that the contribution of non-melonic diagrams will break the interpretation of the IR solution as extremizing the free energy of a collection of free fields (subject to a constraint). Of course, we must still lie at an extremum of $\Gamma_\mathrm{2PI}$. It would therefore be interesting to compute the first correction at subleading order in $N$. In the SYK model, the NLO 2PI vacuum graphs take the form of the periodic ladders with $n \ge 1$ rungs, with or without one twist of the rails \cite{Benedetti:2018goh}; in the tensor models, diagrams containing other couplings enter at NLO, and ladder-like diagrams appear at NNLO \cite{Benedetti:2021wzt}. We note that there are certain subtleties associated to the computation of the 2PI action \cite{Benedetti:2018goh}.
\item Along similar lines, we often find complex solutions for the scaling dimension of some operator -- these are thought to indicate that in the true IR vacuum that operator would condense \cite{Benedetti:2021qyk}. Thus: what happens when we permit the possibility of symmetry breaking, as in \cite{Kim:2019upg}?
\item One obvious generalization that suggests itself is to formally modify the melonic constraint to some generic $f(\{\Delta_\phi\})=0$. This was first explored in \cite{Shen:2023srk}, where a bilocal interaction was used to effectively set the melonic constraint of an SYK-like theory to $q \Delta_\phi = d -\alpha$, for tunable $\alpha$.
\item The so-called higher melonic theories \cite{Gubser:2018yec} fit very neatly into this framework. %
Essentially, we compute $\Ft$ for irreps of $\SO(d+1,1)$ over non-Archimedean fields, and then perform a constrained extremization as usual. Their multiple-field generalizations may then prove rich. This also suggests considering the properties of the melonic-type theories defined over non-compact groups $G$ other than the conformal group; these are then a solvable sector of the $G$-theories proposed in \cite{Gadde:2017sjg}. %
\end{enumerate}

\acknowledgments

The authors are grateful to M\'ark Mezei and John March-Russell for discussions. LFT is supported by a Dalitz Scholarship from the University of Oxford and Wadham College.

For the purpose of open access, the authors have applied a CC BY public copyright licence to any Author Accepted Manuscript (AAM) version arising from this submission.

\appendix
\section{Diagrammatic proof of constrained \FtextOrPDF-extremization} \label{app:diagrammaticProof}

In the following, we demonstrate how $\Ft$-extremization can be recovered by a more standard Feynman-diagrammatic calculation of the two-point Schwinger-Dyson equations.

\subsection{Conformal two-point functions and conventions}

The inverse of a conformal two-point function $\twoPt_\phi$, if it exists, is defined by
\begin{equation}\begin{aligned}
\int \dd^d x \, \twoPt_\phi (w-x)\indices{_{\mu_1 \cdots \mu_s}^{\nu_1 \cdots \nu_s}} \, [\twoPt_\phi^{-1}(x-y)]\indices{_{\nu_1 \cdots \nu_s}^{\sigma_1 \cdots \sigma_s}} &\equiv \delta^d(w-y) \, \hat{\delta}\indices{_{\mu_1\cdots \mu_s}^{\sigma_1 \cdots \sigma_s}},
\end{aligned}\end{equation} 
which we write in the shorthand form
\begin{equation}\begin{aligned}
\int \dd^d x \, \twoPt_\phi (w-x) \cdot \twoPt_\phi^{-1}(x-y) &\equiv \delta^d(w-y)\, \mathbb{I}_{\rho_{\phi}},
\end{aligned}\end{equation} 
where $\mathbb{I}_{\rho_{\phi}} \equiv\hat{\delta}\indices{_{\mu_1\cdots \mu_s}^{\sigma_1 \cdots \sigma_s}}$ is the appropriate identity for $\rho_{\phi}$. As usual, all $\Gglobal$ indices are suppressed.

Now define the unique shadow field $\tilde\phi$ to be the field with scaling dimension $\tilde{\Delta}_\phi \equiv d-\Delta_\phi$, transforming in the \textit{reflected representation}\footnote{The reflected representation of $\rho$ is defined by $\rho^R(g) \equiv \rho(RgR^{-1})$, where $R \in \gO(d)$ is a reflection in any direction. In odd dimensions, $\rho^R$ and $\rho$ are equivalent, because $\gO(d)$ factorizes into $\SO(d) \times \mathbb{Z}_2$ as $-1$ is a reflection matrix.  In even dimensions, to obtain $\rho^R$, we swap the weights associated to the two spinor representations.} $\rho_{\phi}^{\mathrm{ref}}$. The operator $\tilde{\phi}^\dagger$ then has scaling dimension $d-\Delta$ and $\SO(d)$ representation $\rho^\star$ (the dual of $\rho$). The inverse propagator and the shadow field propagator transform identically under conformal symmetry. Since both are unique, we must have%
\begin{equation}\begin{aligned}
\twoPt_\phi^{-1}(x-y) &=\frac{1}{\cN_{\phi}} \twoPt_{\tilde\phi}(x-y), %
\label{eq:inverseIsShadow}
\end{aligned}\end{equation} 
for some $\cN_\phi$; this must be a purely representation-theoretical quantity (since $D_\phi$s are unit normalised) that can be calculated explicitly by taking the inverse in momentum space. This result is a generic identity for CFTs, provided that $\cN_{\phi} \neq \infty,0$, which occurs for operators transforming in the exceptional series of conformal group representations. $\cN_\phi$ can be found using \cite{Karateev:2018oml} %
\begin{equation}\begin{aligned}
(-1)^{\mathrm{F}_\Psi} \frac{\dim_{\mathbb{R}}(\rho_\phi)}{\cN_\phi} = \Omega_d \, \mu(\phi),
\end{aligned}\end{equation} 
where $\Omega_d=2^d \vol\SO(d)$ is a constant that drops out of all computations, $(-1)^{\mathrm{F}_\psi}$ just gives a minus sign for fermionic reps, %
and $\mu(\phi)$ is the Plancherel measure of $\SO(d+1,1)$ for the conformal representation of $\phi$. For comparison with the main text, we note that $\Ft$ for a generalized free field is defined by
\begin{equation}\begin{aligned} \label{eq:FtEqIntMu}
\Ft(\Delta_\phi, \rho_\phi) = \frac{\pi^{d+1}}{\Gamma(d+1)}\int_{\frac{d}{2}}^{\Delta_\phi} \dd \Delta' \, \Omega_d \, \mu(\phi|_{\Delta'}),
\end{aligned}\end{equation} 
where for a scalar field, $\cN_\phi^{-1} = c(\Delta)c(d-\Delta)$.

In the scalar case, these correspond to the GFFs with Lagrangian $\phi (-\partial^2)^{k} \phi$ for $k\in \mathbb{N}_0$ -- in particular,  we have that $\cN_\phi =\infty$ for the free scalar.

\subsection{Two-point Schwinger-Dyson equations}

As is usual for conformal field theories, we will work in position space, where the two-point functions $G_\phi=\cZ_\phi D_\phi$ (defined in \eqref{eq:conformalG}) satisfy the usual Schwinger-Dyson equation %
\begin{subequations}
  \begin{align}%
\vcenter{\hbox{\begin{tikzpicture}
  \begin{feynman}[every blob={/tikz/fill=gray!30}]
    \vertex[small,rectangle, draw=black,fill=gray!30]  (m) at (0,0) {$\mathcal{Z}_\phi D_\phi$};
    \vertex (a) at (-1,0) ;
    \vertex (b) at ( 1,0);
    \diagram* {
      (a) --[scalar] (m) --[scalar] (b),
      };
  \end{feynman}
\end{tikzpicture}}}
\quad &= \quad \vcenter{\hbox{\begin{tikzpicture}
  \begin{feynman}
  \vertex[small, rectangle, draw=black] (m) at (0,0) {$C^{\text{free}}_\phi$};
    \vertex (a) at (-1,0) ;
    \vertex (b) at ( 1,0);
    \diagram* {
      (a) --[scalar] (m) --[scalar] (b),
      };
  \end{feynman}
\end{tikzpicture}}} \quad+ \quad
\vcenter{\hbox{\begin{tikzpicture}
    \begin{feynman}
    \vertex (a) at (-1,0) ;
     \vertex[small, rectangle,draw=black] (m) at (0,0) {$C^{\text{free}}_\phi$};
   \vertex[large,blob,fill=gray!30] (centreblob2) at (1.5,0) {$\Pi_\phi$};
   \vertex[small,rectangle, draw=black,fill=gray!30] (end) at (3,0) {$\mathcal{Z}_\phi D_\phi$};
    \vertex (e) at (4,0);
    \diagram* {
      {[edges={scalar}] (a) -- (m)  -- (centreblob2) -- (end) -- (e)}
      };
    \end{feynman}
  \end{tikzpicture}}}\\
\cZ \twoPt_\phi &= C^{\text{free}}_\phi + C^{\text{free}}_\phi \star \Pi_\phi \star \cZ_\phi \twoPt_\phi.
\end{align}\end{subequations}
Here, $C^{\text{free}}_\phi(x,y)$ is the bare propagator of the field, assumed also to be conformal with scaling dimension $\Delta_\phi^\mathrm{free}$; $\Pi_\phi(x,y)$ is the 1PI self-energy for the field $\phi$; and $\star$ indicates convolution of these two-index objects.
Convolving with the inverses, we find 
\begin{equation}\begin{aligned} \label{eq:inversesRelationship}
[C_\phi^{\text{free}}]^{-1} = \frac{1}{\cZ_\phi} [\twoPt_{\phi}]^{-1} + \Pi_\phi.
\end{aligned}\end{equation} 
We assume that the free propagator drops out. For an IR or UV CFT, this means that each field must satisfy
\begin{equation}\begin{aligned} \label{eq:UVIRscaling}
\text{IR: }\Delta_\phi > \Delta_\phi^{\text{free}}; \qquad \text{UV: }\Delta_\phi < \Delta_\phi^{\text{free}}.
\end{aligned}\end{equation} 
The case of equality, $\Delta_\phi = \Delta_\phi^{\text{free}}$, leads to the long-range melonic models \cite{Gross:2017vhb,Benedetti:2019rja,Benedetti:2020rrq,Benedetti:2021wzt,Shen:2023srk}. %
Note that for canonical free field kinetic terms, any UV CFT must violate the unitarity bounds.

\subsection{Melonic theories}

We will specify a melonic-type theory with schematic interaction Lagrangian\footnote{Note that $\tilde{g}_m$ is the coupling constant with the conventional Feynman-diagrammatic normalising factor associated to a melon. This differs from the coupling normalisation commonly used for SYK.}
\begin{equation}\begin{aligned}
\sum_{m}^{n_m} \tilde{g}_m \prod_\Phi \frac{\Phi^{q^m_\Phi}}{q^m_\Phi!}.
\end{aligned}\end{equation}  
For convenience, we discuss only bosonic fields. %
As discussed in the main text, we assume that there is an unspecified underlying mechanism (typically a disorder average or tensorial structure) that enforces the large-$N$ melonic dominance.
It is standard that the self-energy $\Pi_\phi$ of each field can then be resummed to
\begin{equation}\begin{aligned}
\Pi_\phi(x,y) =\sum_m \vcenter{\hbox{\begin{tikzpicture}[anchor=base,baseline]
  \pgfmathsetmacro{\L}{4}
  \coordinate (L) at (-\L,0);
  \coordinate (R) at (\L,0);
  
  \draw[black, decorate, decoration={snake, amplitude=.4mm, segment length=2mm}]  (L) to[bend left=90,looseness=1.25] (R);
  \draw[black, decorate, decoration={snake, amplitude=.4mm, segment length=2mm}]  (L) to[bend left=75,looseness=1.20] (R);
  \draw[loosely dotted, thick] (0, {0.65*\L}) to (0, {0.55*\L});
  \node at (0.4, {(0.58*\L)}) {$q^m_{\Phi_i}$};
  \draw[black, decorate, decoration={snake, amplitude=.4mm, segment length=2mm}]  (L) to[bend left=60] (R);
  
  \draw[black, thick] (L) to[bend left=50] (R);
  \draw[black, thick] (L) to[bend left=40, looseness=1] (R);
  \draw[loosely dotted, thick] (0, {0.32*\L}) to (0, {0.22*\L});
  \node at (0.8, {(0.25*\L)}) {$(q^m_{\phi} -1)$};
  \draw[black, thick] (L) to[bend left=25, looseness=0.7] (R);
  
  \node at (-\L-0.3, 0) {$x$};
  \draw[loosely dotted, thick] (0, {0.09*\L}) to (0, {-0.09*\L});
  \node at (\L+0.3, 0) {$y$};
  
  \draw[black, dashed] (L) to[bend right=50] (R);
  \draw[black, dashed] (L) to[bend right=40, looseness=1] (R);
  \draw[loosely dotted, thick] (0, {-0.23*\L}) to (0, {-0.33*\L});
  \node at (0.4, {-(0.29*\L)}) {$q^m_{\Phi_k}$};
  \draw[black, dashed] (L) to[bend right=25, looseness=0.7] (R);
\end{tikzpicture}}},
\end{aligned}\end{equation} 
where the propagator on each leg is the full resummed propagator $D_\Phi$. Each diagram has symmetry factor $\prod_\Phi q^m_\Phi!/q^m_\Phi$, so in the scaling limit \eqref{eq:inversesRelationship} becomes
\begin{subequations}
\begin{equation}\begin{aligned}
\frac{-1}{\cZ_\phi} [\twoPt_\phi(x-y)]^{-1} =\frac{1}{\dim(\rhoext_\phi)} \sum_{m} \tilde{\mathfrak{g}}_m q^m_\phi [\cZ_\phi \twoPt_\phi (x-y)]^{q^m_\phi -1} \, \prod_{\Phi \neq \phi} [\cZ_\Phi \twoPt_{\Phi}(x-y)]^{q^m_{\Phi}},
\end{aligned}\end{equation} 
where
\begin{equation}\begin{aligned}
\tilde{\mathfrak{g}}_m = \frac{\tilde{g}_m^2}{\prod_{\Phi} q^m_\Phi!}.
\end{aligned}\end{equation} \label{eq:2ptSDEfull}
\end{subequations}
Dimensional analysis of \eqref{eq:2ptSDEfull} tells us immediately that the continuous data of this melonic theory, being the scaling dimensions, are forced to obey the following equality for each melon $m$:
\begin{equation}\begin{aligned}
\sum_{\Phi} q^m_{\Phi} \Delta_\Phi = d. \label{eq:dimSumRuleGeneral}
\end{aligned}\end{equation} 
By \eqref{eq:inverseIsShadow}, the right-hand side must transform in the shadow representation of the field $\Phi$, that is $\rhoext_{\tilde{\Phi}}$. For this reason, we do not need to keep track of the Lorentz indices: the various contractions must end up giving an identity $\id_s$ on the right-hand side, and so only contribute a factor of $\dim(\rho_\Phi)$ inside $\dim(\rhoext_{\psi})$. Likewise, since the symmetry group is assumed unbroken, the $G$ indices marshal themselves into a $\id_R$.

As before, the $D_\phi$s are unit normalised, and therefore we must have
\begin{equation}\begin{aligned}
[\twoPt_\phi (x-y)]^{q^m_\phi -1} \, \prod_{\Phi \neq \phi} [\twoPt_{\Phi}(x-y)]^{q^m_{\Phi}} & = \twoPt_{\tilde\phi}(x-y).
\end{aligned}\end{equation} 
Plugging that in to \eqref{eq:2ptSDEfull}, we find
\begin{equation}\begin{aligned}
[\cZ_\phi \twoPt_\phi(x-y)]^{-1} &= -\frac{1}{\dim \rhoext_\phi} \sum_{m} q^m_\phi \mathfrak{g}_m \twoPt_{\tilde\phi}(x-y),
\end{aligned}\end{equation} 
where we have defined a renormalized coupling constant for each melon
\begin{equation}\begin{aligned}
\mathfrak{g}_m \equiv \tilde{\mathfrak{g}}_m \left(\prod_{\Phi}^{\text{melon}} Z_\Phi^{q_\Phi}\right) = g_m^2 \prod_\Phi \frac{Z_\Phi^{q_\Phi^m}}{q^m_\Phi !}.
\end{aligned}\end{equation} 
Then using the identity \eqref{eq:inverseIsShadow}, for each field $\phi$ we obtain
\begin{equation}\begin{aligned}
\frac{\dim\rhoext_\phi}{\cN_\phi} =- \sum_{m} q^m_\phi  \mathfrak{g}_m. \label{eq:otherGeneralRule}
\end{aligned}\end{equation} 
Thus, if we have $n_m$ melons and $n_f$ fields $\{\phi\}$ in $\SO(d)\times G$ representations $\rhoext_{\phi}$, we have $n_m$ equations from \cref{eq:dimSumRuleGeneral} and $n_f$ equations from \cref{eq:otherGeneralRule}. Therefore, generically we can find a solution for the unknown $\mathfrak{g}_m$s and $\Delta_{\phi}$s. 

Recall that we required $\Delta_{\phi} > \Delta^{\text{free}}_{\phi}$ in order to obtain consistent IR scaling. This picks out a polyhedron of allowed scaling dimensions in $\mathbb{R}^{n_f}$.
However, we can always tune the free scaling dimension of the fields by modifying the kinetic terms; and so we ignore this condition. %
Using \eqref{eq:FtEqIntMu}, we can eliminate $\cN_\phi$ in favour of $\dv{\Ft_\phi}{\Delta_\phi}$, and so derive:
\begin{subequations}
\begin{empheq}[box=\widefbox]{align}
\text{Given a set of melonic data: }& \{m: \,\, \prod_\Phi \Phi^{q^m_\Phi} \}\\
\text{For each melon: }& \sum_{\Phi}^{\text{melon}} q^m_{\Phi} \Delta_\Phi = d\\
\text{For each field $\phi$: }&  \dv{\Ft_\phi(\Delta_\phi)}{\Delta_\phi} =- \frac{\pi^{d+1}}{\Gamma(d+1)} \sum_{m} q^m_\phi  \mathfrak{g}_m \label{eq:dimrhoOverN}
\end{empheq}\label{eq:melonicsummary}
\end{subequations}
These equations give a complete solution for the conformal melonic limit, assuming appropriate IR (UV) scaling \eqref{eq:UVIRscaling}. We have assumed here that all melons are IR-relevant; if they are not, $\sum_{\Phi} q^m_\Phi \Delta_\Phi > d$, then they will drop out in the IR. These equations \eqref{eq:melonicsummary} are manifestly recoverable from the $\Ft$-extremization shown in the main text.

\bibliographystyle{JHEP} %
{\raggedright  %
\bibliography{references}    %
}              %

\end{document}